\documentclass[twocolumn,tradiabstract]{aa} 

\usepackage{natbib}
\usepackage{graphicx}

\def\specchar#1{{\sc #1}}
\def\FeI{\mbox{Fe\,\specchar{i}}}
\def\FeII{\mbox{Fe\,\specchar{ii}}}
\def\HI{\mbox{H\,\specchar{i}}}
\def\HeI{\mbox{He\,\specchar{i}}}
\def\HeII{\mbox{He\,\specchar{ii}}}
\def\CaII{\mbox{Ca\,\specchar{ii}}}
\def\Halpha{\mbox{H$\alpha$}}

\newcommand\mancha{\textsc{Mancha3D~}}

\begin{document}

\title{Two-fluid simulations of waves in the solar chromosphere I: numerical code  verification.}
   
\author{B. Popescu Braileanu\inst{1,2}\thanks{ \email{bpopescu@iac.es}}
          \and
          V. S. Lukin\inst{3}
          \and
          E. Khomenko\inst{1,2}
          \and
          \'A. de Vicente\inst{1,2}}

\titlerunning{Two-fluid simulations of waves in the solar chromosphere}
\authorrunning{B. Popescu Braileanu et al.}

\institute{Instituto de Astrof\'{\i}sica de Canarias, 38205 La Laguna, Tenerife, Spain
\and Departamento de Astrof\'{\i}sica, Universidad de La Laguna, 38205, La Laguna, Tenerife, Spain
\and National Science Foundation\thanks{Any opinion, findings, and conclusions or recommendations expressed in this material are those of the authors and do not necessarily reflect the views of the National Science Foundation.}, Alexandria, VA, 22306, USA}

\date{Received:  August 29, 2018; Accepted: May 7, 2019}
 
\abstract{Solar chromosphere consists of a partially ionized plasma, which makes modeling the solar chromosphere a particularly challenging numerical task.  Here we numerically model chromospheric waves using a two-fluid approach with a newly developed numerical code.  The code solves two-fluid equations of conservation of mass, momentum and energy, together with the induction equation, for the case of the purely hydrogen plasma with collisional coupling between the charged and neutral fluid components. 
The implementation of a semi-implicit algorithm allows us to overcome the numerical stability constraints due to the stiff collisional terms.  We test the code against analytical solutions of acoustic and Alfv\'en wave propagation in uniform medium in several regimes of collisional coupling.The results of our simulations are consistent with the analytical estimates, and with other results described in the literature.  In the limit of a large collisional frequency, the waves propagate with a common speed of a single fluid.  In the other limit of a vanishingly small collisional frequency, the Alfv\'en waves propagate with an Alfv\'en speed of the charged fluid only,  while the perturbation in neutral fluid is very small.  The acoustic waves in these limits propagate with the sound speed corresponding to either the charges or the neutrals,  while the perturbation in the other fluid component is very small.  Otherwise, when the collision frequency is similar to the real part of the wave frequency, the interaction between charges and neutrals through momentum transfer collisions cause alterations of the waves 
frequencies and damping of the wave amplitudes. }

\keywords{Sun: chromosphere -- Sun: waves -- Sun: magnetic field -- Sun: numerical simulations}

\maketitle
%

\section{Introduction}

Solar chromosphere is a transition layer where the properties of plasma change abruptly from gas pressure dominated (as in the photosphere) to magnetic field  dominated (in corona),  and where the collisional coupling of the plasma weakens with height.  Several aspects of chromospheric physics make this region one of the most complex for understanding of the solar atmosphere.

It is well known that the chromosphere is not in local thermodynamic equilibrium.  \cite{1962Athay} measured and compared intensities in strong emission lines  of H, \HeI, \HeII, \CaII,  and the deduced ratios of the populations of various levels indicated a significant departure from a Boltzmann distribution of the levels studied.   In the high chromosphere the ratio of neutral atoms to ions exceeds the LTE values by about  10$^7$-10$^8$ in the case of \HeI\ and by about  10$^6$-10$^7$ in the case of hydrogen,  and there is evidence that significant  departures from LTE extend well  into the photospheric layers.   The formation of many photospheric lines is well described using LTE when radiation is fully coupled to  the local physical conditions.  However,  in the chromosphere when collisional rates are lower,  the radiation begins to decouple from the local physical conditions, invalidating the LTE assumption.

Plasma in the chromosphere is only partially ionized.  As the density decreases with height, the collision frequencies between different particle species decrease and the magnetic field begins to dominate the dynamics.  The neutral species do not feel the magnetic field directly and this causes a partial decoupling between charges and neutrals  when the collisional time scales between ionized and neutral  atomic species become equal or larger than the  hydrodynamic time scale. Therefore, classical MHD approach is not the best approximation to treat the plasma processes in the chromosphere.   While realistic photospheric models based on the MHD approximation  provide results that are practically indistinguishable from observations even in strongly magnetized regions \citep[e.g.,][]{2009Rempel}, this is not the case for the chromosphere \citep[see e.g. the discussion in][]{Leenaarts2010}. 
This is so because observations do not reach scales where ion-neutral effects can be detected directly in plasma dynamics,
\citep[see e.g. Figure 1 in][]{2014Khomenko} in the photosphere. However, in the chromosphere, these scales become larger
and may be possible to reach in some dedicated observations. Indirect ion-neutral decoupling effects, such as heating, are also more pronounced
in the chromosphere.
 
A suitable alternative to the MHD approach is a two-fluid plasma-neutrals model, numerically implemented in this work.  This kind of a model is  most appropriate for high density plasmas such as in the chromosphere, where kinetic models cannot be efficiently applied. 
Kinetic models have been successfully applied to the corona and the solar wind, where much lower particle density makes the problem more computationally feasible and the kinetic description more necessary and appropriate.
Multi-fluid models preserve a number of important  physical effects coming from weak collisional coupling, still allowing for an efficient implementation, as any fluid-based description. While the motivation for multi-fluid modeling comes mainly from theoretical considerations,  in the recent years there have been attempts of direct observational detection of multi-fluid effects in the solar plasma through measurements of differences  in ion and neutral velocities caused by loss of collisional coupling. By measuring the Doppler shift in ion \FeII\ and neutral \FeI\ lines, simultaneously over the same volume of plasma,   differences between ion and neutral velocities of the Evershed flow have been found in sunspot penumbra as deep as  the photosphere by  \cite{2015Khomenko}.  Similarly, \cite{2016Khomenko} showed that non-negligible differences in \HeI\ and \CaII\ velocities  exist in solar prominences when observed with sufficiently high cadence. While the measured neutral and ion velocities were the same in most of the spatial locations most of the time,  they were observed to decouple in the presence of  strong spatial and temporal gradients, such as wave fronts.  However,   \cite{2017Anan} concluded that the observed small differences in velocities of  \HI\ 397 nm,  \HI\ 434 nm, \CaII\ 397 nm, and \CaII\ 854 where indications of motion of different components in the prominence  along the line of sight, rather than indications of decoupling.  Detection of the decoupling effects directly would require highest spatial and temporal resolution,  apart from a careful selection of spectral lines for the analysis.  More observational studies in the future will be needed to confirm or discard the detection of the decoupling effects.

Nonetheless, there  exist other indirect indications of decoupling such as the those observed by \cite{2007Gilbert}.  These authors have compared \HeI\ 1083 nm and \Halpha\ data in multiple solar prominences in different phases of their lifecycle,  and were able to detect the drainage effects across prominence magnetic field with different time scales for helium and hydrogen atoms. Such drainage was predicted earlier by \cite{2002Gilbert} from simplified semi-analytical calculations. Another indirect evidence of ion-neutral decoupling  was reported by \cite{2011delaCruz}  who deduced misalignment in the visible direction of chromospheric fibrils and measured magnetic field vector. They noted most of the fibrils aligned with the magnetic field, although there are a few noteworthy cases where significant misalignments occur, well beyond the observational uncertainty.

Misalignment of chromospheric fibrils and magnetic field was confirmed by \cite{2016Sykora}  who used advanced radiative magnetohydrodynamic simulations, including the effects of ion-neutral interactions in the single-fluid approximation (via ambipolar diffusion). These authors have shown that the magnetic field is indeed often not well  aligned with chromospheric features. The misalignment occurs where the ambipolar diffusion is large, i.e., ions and neutral populations decouple as the ion-neutral collision frequency drops,  allowing the field to slip through the neutral population. The conditions for misalignment also require currents perpendicular to the field to be strong, and thermodynamic time scales to be longer than or similar to those of ambipolar diffusion. 

The vast majority of works studies of ion-neutral effects in solar plasmas address propagation of different types of waves and the development of  instabilities in chromospheric and prominence plasma conditions.  \citet{2011Zaq} and \citet{2013Soler} studied theoretically  the Alfv\'en waves propagation in a uniform medium composed by protons and hydrogen atoms which interact by collisions.  Refinements of this theory were introduced by \citet{2011Zaq2} by adding helium atoms, by \citet{2013Zaq} by considering a nonuniform gravitationally stratified atmosphere, and by  \citet{2013Soler2} by applying a flux tube model. It was found that in situations when the neutral-ion collisional frequency is much lower than the wave frequency, the propagation properties of the waves only depend on the properties of the ionized fluid. These properties change with height due to stratification, and the Alfv\'en wave may become evanescent in some regions of the solar atmosphere. The waves can be damped by neutral-ion collisions, and the existence of neutral helium atoms, alongside with neutral hydrogen, significantly enhances the damping.  The damping is most efficient when the wave frequency and the collisional frequency are of the same order of magnitude.  The overall conclusion from these studies is that  the single-fluid approach is suitable for dealing with slow processes in partially ionized plasmas, which happen on time scales longer than the collisional time scale. This approximation fails for time scales  comparable to or shorter than the characteristic time scales for interaction between the charged and neutral particle populations. 

Only recently, several numerical simulations of solar phenomena using a two-fluid approach have been reported.  However, their realism is rather limited. \cite{2017Maneva} modeled  magneto-acoustic wave propagation in the solar  stratified chromosphere  including effects of impact ionization and radiative recombination. These authors found numerous difficulties in constructing an equilibrium model atmosphere for the wave propagation,  and in interpreting the results of their simulations in comparisons to more standard, single fluid models. Magnetic reconnection has been studied using a two fluid model by \cite{2008Sakai, 2012Lukin, 2013Lukin, 2014Laguna, 2016Hillier}. Important decoupling between the flows of neutrals and ions where observed in their simulations close to a reconnection site with  considerable differences between inflows and outflows.  \cite{Ni2018}  also found that even in the absence of flow decoupling, dynamics of the ionization and recombination processes can have important consequences for the  thermodynamics and observable properties of plasma at chromospheric reconnection sites.  \cite{2017Gomez} used a five-fluid model with three ionized and two neutral components, which takes into consideration  Hall current and Ohm's  diffusion in addition to the friction due to collisions between different species.   They apply their model to study the wave propagation in homogeneous plasmas composed of hydrogen and helium, similar to  analytical models described above, but allowing time evolution of the ionization fraction.  They confirmed the theoretical result that the inclusion of  neutral components in the plasma modifies the oscillation periods of the  low-frequency Alfv\'en waves and that collisions produce damping of the perturbations,   the damping being more efficient when the collisional frequency is of the order of the oscillation frequency. 

From all of the above it is evident that numerical modeling of solar chromospheric plasma including multi-fluid effects is a promising approach,  and that a significant effort is required in this direction.  Both previous theoretical and observational results suggest the need for such modeling. In the current work we describe the results from the newly developed two-fluid code for purely hydrogen plasma which  takes into account the effects of  elastic and inelastic interactions between electrons, ions, and neutral atoms. In the current paper, which is the first paper in our study of chromospheric waves and shocks within a multi-fluid approximation, we verify the code against the known analytical solutions for the wave propagation in a homogeneous plasma,  and provide results on the numerical convergence of the implemented numerical scheme. 
In section \ref{sec:test_grav} we describe a test of a fast wave in a gravitationally stratified atmosphere, and test the numerical solution
against the  analytical solution.
The full equations  are being provided in this first paper, describing the upgraded code, in order to enable reproducibility, and for future reference in the follow-on papers. 
In the second paper of the series \citep{Popescu2018b}, we apply the code to study the non-linear behavior of chromospheric shocks under two-fluid condition.

\section{Numerical method}

A multicomponent plasma evolution can be described by equations of continuity, momentum and energy of its components which are obtained  
via momenta (0th, 1st, and 2nd moment) of the Boltzman equation.  
Typically, ions(i), electrons(e) and neutrals(n) are considered. When treated numerically by an explicit code,  
the equation of evolution of electrons imposes very small time steps and therefore is numerically  restrictive. 
Because of 
the large ion-to-electron mass ratio, the electron propagation speed, sound speed and Alfven velocity are also large relative to the corresponding bulk plasma parameters; this restricts the time step in an explicit scheme.
However, for the same reason, the electron mass effects can generally be neglected from the electron equation of motion.  This relaxes some of the numerical constraints and results in a generalized Ohm's law used to compute evolution of the electric field. The following assumptions are made in the set of two-fluid equations implemented in this work:
\begin{itemize}
\item Purely hydrogen plasma. This condition implies than only singly-ionized ions are present and that charge neutrality is fulfilled on scales above the Debye radius scale, as usual. Charge exchange reactions are introduced via  elastic collisions, by defining
a charge-exchange collisional parameter $\alpha_{cx}$, added to the elastic collisional parameter $\alpha$. The expression of this term is given in Equation (\ref{eq:alpha_cx}), in Appendix \ref{app:collTerms}. For such plasma, a constant mean molecular weight for neutrals is $\mu_n$=1 g/mole and a constant mean molecular weight for charges is $\mu_c$=0.5 g/mole,  where the sub-index ``c'' stands for ``charges''.
\item Same temperatures of ions and electrons, i.e., $T_c = T_e = T_i$. Neutrals are allowed to have different temperatures than charges, i.e. generally $T_n \ne T_c$.
\item Center of mass velocity of charges is essentially ion velocity, i.e., $\vec{u}_c = \vec{u}_i$. This condition implies that $\vec{u}_e = \vec{u}_c - \frac{\vec{J}}{en_e}$.
\end{itemize}
%
\subsection{Equations}
%
We solve continuity, momentum and energy conservation equations written separately for charges (sub-index ``c'') and neutrals (sub-index ``n''). The interaction between neutrals and charges is introduced via the different kinds of collisional terms, as follows. \\

\noindent \textbf{Continuity equations:}
\begin{align}  \label{eq:continuity}
\frac{\partial \rho_n}{\partial t} + \nabla \cdot (\rho_n\vec{u}_n) &= S_n, \\ \nonumber
\frac{\partial \rho_c}{\partial t} + \nabla \cdot (\rho_c\vec{u}_c) &= -S_n,
\end{align}
with: 
\begin{equation} \label{eq:s}
S_n = \rho_c \Gamma^{\rm rec} - \rho_n\Gamma^{\rm ion},
\end{equation}
being the ionization/recombination collisional source term. The variables $\Gamma^{\rm ion}$ and $\Gamma^{\rm rec}$ are ionization and recombination rates.  They are defined in the Appendix \ref{app:collTerms}. The variables $\rho_c$ and $\rho_n$ are mass density of charges and neutrals, and $\vec{u}_c$ and $\vec{u}_n$ are their center of mass velocities. \\

\noindent \textbf{Momentum equations:}

\begin{align} \label{eq:momentum}
&\frac{\partial (\rho_n\vec{u_n})}{\partial t} + \nabla \cdot (\rho_n\vec{u_n} \otimes \vec{u_n} +{\bf\hat{p}_n} )
= -\rho_n\vec{g} +\vec{R}_n, \\ \nonumber
&\frac{\partial (\rho_c\vec{u_c})}{\partial t} + \nabla \cdot (\rho_c\vec{u}_c\otimes\vec{u}_c + {\bf\hat{p}_c} )
=\vec{J}\times\vec{B} - \rho_c\vec{g}  -\vec{R}_n\,.
\end{align}
\noindent The elastic collisions between ions and neutrals and between electrons and neutrals can be combined into a single collision frequency
between charges and neutrals. By defining the collisional parameter:
\begin{equation} \label{eq:alpha}
\alpha=\frac{\rho_{e} \nu_{en} + \rho_{i} \nu_{in}}{\rho_{n} \rho_{c}},
\end{equation}
the effective collision frequency between charges and neutrals, and between neutrals and charges becomes, respectively: 
\begin{eqnarray} \label{eq:eff_coll}
\nu_{cn} &= \alpha \rho_n, \nonumber \\
\nu_{nc} &= \alpha \rho_c \,.
\end{eqnarray}
Here, $\nu_{en}$ and  $\nu_{in}$ are the electron-neutral and ion-neutral collision frequency, respectively, with expressions 
given in Equation \ref{eq:nu_el}, in  Appendix \ref{app:collTerms}.
If we take into account the charge-exchange reactions, $\alpha_{\rm cx}$, as defined 
in Equation (\ref{eq:alpha_cx}), should be added to the elastic collision parameter $\alpha$.
The momentum collisional term $\vec{R}_n$ can be written as:
\begin{equation} \label{eq:r}
\vec{R}_n = \rho_c \vec{u}_c \Gamma^{\rm rec}  - \rho_n \vec{u}_n \Gamma^{\rm ion} 
+ \alpha \rho_n \rho_c (\vec{u}_c - \vec{u}_n).
\end{equation}
Here ${\bf\hat{p}_c}$ and ${\bf\hat{p}_n}$ are pressure tensors for charged and neutral species, $\vec{B}$ is magnetic field, $\vec{J}$ is current density and $\vec{R}_n$ is momentum collisional exchange term.  The latter takes into account momentum gains/losses due to elastic collisions, with collisional frequencies between ions, electrons and neutrals, $\nu_{in}$ and $\nu_{en}$, defined in the Appendix \ref{app:collTerms}.  It also takes into account momentum gains/losses due to ionization and recombination (the terms proportional to the ionization/recombination rates $\Gamma$).

We consider isotropic pressure, but we include viscosity in the pressure tensor:
\begin{equation} 
{\bf\hat{p}}_\alpha =  p_\alpha \mathbb{I} - {\bf\hat{\tau}}_\alpha,
\end{equation}
where the elements of  the viscosity tensor are:
\begin{equation} \label{eq:visc}
{\tau_\alpha}_{ij} =  \xi_\alpha \left ( \frac{\partial {u_\alpha}_i}{\partial x_j} + \frac{\partial {u_\alpha}_j}{\partial x_i} \right ) \,.
\end{equation}
\noindent   The expressions used for  the viscosity coefficient for specie $\alpha$, denoted by the symbol $\xi_\alpha$ are given  in 
Equation (\ref{eq:visc_th_coef}), in Appendix \ref{app:collTerms}.

The gravitational acceleration $\vec{g}$ is oriented in the  negative  z direction.\\

\noindent \textbf{Energy conservation equations:}

\begin{eqnarray} \label{eq:energy}
\frac{\partial}{\partial t} \bigl( e_n +\frac{1}{2}\rho_n u_n^2 \bigr) &+& \nabla \cdot \bigl( \vec{u}_n (e_n + 
\frac{1}{2}\rho_n u_n^2 )  + {\bf\hat{p}_n} \cdot \vec{u}_n \nonumber \\ - \ K_n \vec{\nabla} T_n\bigr ) &=& -\rho_n\vec{u}_n \cdot \vec{g}  + M_n,  \nonumber \\
\frac{\partial}{\partial t} \bigl(  e_c+\frac{1}{2}\rho_c u_c^2 \bigr) &+&  \nabla \cdot \bigl( \vec{u}_c (e_c + 
\frac{1}{2}\rho_c u_c^2 )  + {\bf\hat{p}_c} \cdot \vec{u}_c \nonumber \\ - K_c \vec{\nabla} T_c  \bigr) &=& -\rho_c\vec{u}_c \cdot \vec{g} + \vec{J} \cdot \vec{E} -M_n,
\end{eqnarray}
\noindent with:
\begin{eqnarray} \label{eq:m}
M_n &=& \left ( \frac{1}{2} \Gamma^{\rm rec} \rho_c u_c^2  - \frac{1}{2}\rho_n u_n^2 \Gamma^{\rm ion} \right )  
\nonumber \\ &+& \frac{1}{\gamma-1} \frac{k_B}{m_n} \left ( \rho_c T_c \Gamma^{\rm rec} - \rho_n T_n \Gamma^{\rm ion} \right) \nonumber \\ 
&+& \frac{1}{2} ({u_c}^2 - {u_n}^2) \alpha \rho_c \rho_n  \nonumber \\
&+&\frac{1}{\gamma-1} \frac{k_B}{m_n}(T_c - T_n)\alpha \rho_c \rho_n\,.
\end{eqnarray} 

\noindent In these equations, $e_c$ and $e_n$ are charges and neutral internal energy, and $T_c$ and $T_n$ are the corresponding temperatures, $K_c$ and $K_n$ are thermal conductivity coefficients,  whose expresions are shown in Equation (\ref{eq:visc_th_coef}).
From the four terms of the expression of $M_n$, the first two are related to ionization/recombination processes, and the
last two terms are related to elastic collisions. 
Terms containing temperatures, the second and the fourth term,  represent the thermal exchange between charges and 
neutrals due to inelastic and elastic collisions, respectively.
Because the energy equations evolve kinetic and internal energy, the $M_n$ terms include kinetic energy gains/losses due to
the change in  the number of particles, and to the work done by the collisional terms.
One can write equations of evolution of internal energy alone:
\begin{align}
\frac{\partial e_n}{\partial t}  +  \nabla \cdot  ( \vec{u}_n e_n  - K_n \vec{\nabla} T_n )  +p_n \nabla \cdot \vec{u}_n - 
\hat{\tau}_n : \vec{\nabla}\vec{u}_n &= M_n', \nonumber  \\
\frac{\partial e_c}{\partial t}  +  \nabla \cdot ( \vec{u}_c e_c - K_c \vec{\nabla} T_c )  + p_c \nabla \cdot \vec{u}_c 
- \hat{\tau}_c : \vec{\nabla}\vec{u}_c &= \nonumber \\ \vec{J} \cdot \vec{E}_{\rm diff}  + M_c', &
\end{align}
where:
\begin{equation} \nonumber
\vec{E}_{\rm diff} = \vec{E} + \vec{u}_c \times \vec{B} 
\end{equation}
is the non-ideal part of the electric field $\vec{E}$, and 
\begin{equation*}
\hat{\tau_\alpha} : \vec{\nabla}\vec{u} =  \xi_\alpha \sum_{i<j}{ \left( \frac{\partial {u_\alpha}_i}{\partial x_j} + \frac{\partial {u_\alpha}_j}{\partial x_i} \right)^2} 
\end{equation*}
is the heating due to the viscosity, being a positive quantity.
Then, the collisional terms become, for neutrals and charges:
\begin{align} \label{eq:mint}
M_n^\prime =  M_n  - \vec{u}_n \cdot \vec{R}_n &+ \frac{1}{2}u_n^2 S_n =  \frac{1}{2} \Gamma^{\rm rec} \rho_c (\vec{u}_c - \vec{u}_n)^2  \nonumber \\ 
 &+\frac{1}{\gamma-1} \frac{k_B}{m_n} \left ( \rho_c T_c \Gamma^{\rm rec} - \rho_n T_n \Gamma^{\rm ion} \right)   \nonumber \\ 
  &+ \frac{1}{2} ({\vec{u}_c} - {\vec{u}_n})^2 \alpha \rho_c \rho_n \nonumber \\ 
 &+\frac{1}{\gamma-1} \frac{k_B}{m_n}(T_c - T_n)\alpha \rho_c\rho_n,  \nonumber \\
M_c^\prime =  -M_n  + \vec{u}_c \cdot \vec{R}_n &- \frac{1}{2}u_c^2 S_n =  \frac{1}{2} \Gamma^{\rm ion} \rho_n (\vec{u}_c - \vec{u}_n)^2 \nonumber \\ 
 &-\frac{1}{\gamma-1} \frac{k_B}{m_n} \left ( \rho_c T_c \Gamma^{\rm rec} - \rho_n T_n \Gamma^{\rm ion} \right)  \nonumber \\ 
  &+ \frac{1}{2} ({\vec{u}_c} - {\vec{u}_n})^2 \alpha \rho_c \rho_n  \nonumber \\  
 &-\frac{1}{\gamma-1} \frac{k_B}{m_n}(T_c - T_n) \alpha \rho_c \rho_n \,. 
\end{align} 
We observe that the second and the fourth terms remain unchanged, 
however, the first and the third terms  are different, they are positive quantities, so they are related to heating. 
The neutrals and charges are heated because of recombination and ionization processes, respectively (the first term),  and both
species are heated at the same rate by elastic collisions (the third term, equal for neutrals and charges, also called
"frictional heating").
Most expressions of the terms related to collisions (including coefficients of viscosities, 
thermal conductions, and  magnetic resisitivity) are derived by \cite{1965Braginskii}, but they are
also detailed in the descriptions of their models by \cite{2012Meier,2012Lukin,2013Lukin}.

The relation between internal energy and pressure, and between the pressure, the number density and the temperature  are defined by  ideal gas laws:
\begin{align}\label{eq:eos}
e_{c,n} &= p_{c,n}/(\gamma -1), \nonumber \\
p_{c,n} &= n_{c,n} k_B  T_{c,n}.
\end{align}

\noindent \textbf{Ohm's law}:  \\

\noindent In order to obtain the Ohm's law we operate the ions and electrons momentum equation. For the case of purely hydrogen plasma, and by neglecting terms proportional to $m_e/m_i$, ions and electrons inertia terms,  and time variation of currents, see \citep{2014Khomenko, 2018Ballester}, it leads to the following expression: 
\begin{equation} \label{eq:ohm} 
\vec{E} + \vec{u}_c\times{\vec{B}} = \eta_H \vec{J}\times \vec{B}  - \eta_H \vec{\nabla}{p_e}  + 
\eta \vec{J}  - \eta_D(\vec{u}_c - \vec{u}_n),
\end{equation}
where:
\begin{eqnarray} 
\eta_H &=& \frac{1}{en_e}, \nonumber \\
\eta &=& \frac{\rho_e\left (\nu_{en}+\nu_{ei} \right ) }{(en_e)^2}, \nonumber \\
\eta_D &=& \frac{\rho_e(\nu_{en} - \nu_{in})}{en_e}\,.
\end{eqnarray}
\noindent All  four terms on the right side of the above equation are the same as in the single fluid approach, and the
first three of them represent: the Hall term, the Biermann battery term, and the Ohmic term. 
The fourth term, proportional to the difference 
of the  velocities of the two species, is usually neglected in both approximations.
The ambipolar term, which appears in the 
single fluid approximation as a consequence of the charged and neutral fluids having different center-of-mass velocities does not appear in the two-fluid approach.
The derivation of these terms can be found in \cite{2014Khomenko}.
\noindent The induction equation for the evolution of the magnetic field is obtained using Maxwell equations, in a usual way,
\begin{eqnarray} \label{eq:ind}
\frac{\partial\vec{B}}{\partial t} = -\vec{\nabla} \times \vec{E} \,.
\end{eqnarray}
%
\subsection{Numerical implementation}

\subsubsection{\mancha code}

In this work we have extended the \mancha code, a single-fluid code, where the effects of partial ionization had been implemented through a generalized Ohm's law including Ohmic, ambipolar, Hall and Biermann battery terms. The new code solves the two-fluid equations in approximation of purely hydrogen plasma, as described above. Ions and electrons are evolved together through the same continuity, momentum and energy equations, and neutrals are evolved separately. The neutral and charges equations are coupled by collisions. These new developments are described below.

The original \mancha code is described elsewhere \citep{2006ManchaKh, Felipe2010, Khomenko+Collados2012, Pedro2018}.  The code is fully 3D, written in Fortran 90, parallelized with MPI, and the output files are in HDF5 format.  The units used in the equations are SI units.  The variables are split into background and perturbation variables, keeping all non-linear terms.  The viscous forces and the Ohmic dissipation tend to eliminate high frequency oscillations.  However, collisional viscosity:  $\xi_{\rm c,n}$ (Eq. \ref{eq:visc}) and Ohmic:  $\nu$ (Eq. \ref{eq:ohm}) coefficients for the parameters of the solar atmosphere are too small  to effectively damp non-resolved variations of different parameters at the grid cell size.  Therefore, for numerical stability, these coefficients have corresponding numerical analogues, called hyperdiffusive coefficients.  The hyperdiffusive coefficients correspond to physical terms: magnetic Ohmic diffusion, conductivity, viscosity, etc.  The amplitude of the hyperdiffusive coefficients is a complex function of space coordinates and of  gradients of the corresponding variables (for details see  \cite{2005ManchaVo, Felipe2010}).  Since sometimes high numerical diffusion is not desirable, an additional filtering of small wavelengths, following \cite{2007ManchaPa}  is implemented in \mancha. In order to avoid reflections at the boundaries,  \mancha uses the Perfect Matching Layer technique \citep{Felipe2010}.  The spatial discretization scheme is fourth order accurate.  The original temporal discretization scheme was a fourth order explicit Runge-Kutta,  modified to an arbitrary order in the latest version of the code due to implementation of STS and HDS schemes \citep{Pedro2018}. The two-fluid implementation required further modification of the explicit Runge-Kutta integration.  The new implementation is a semi-implicit scheme with up to second order accuracy,  as is shown in the numerical tests of the code described below.

\subsubsection{Numerical scheme}

Since the solar atmosphere is strongly gravitationally stratified, the collisional frequencies change orders of magnitude  from the photosphere to the chromosphere.   In the deep layers, collisional terms are typically very large and dominate over the hydrodynamical time scales.   In the chromosphere this situation may change, since collisional frequencies become significantly smaller.   Therefore, the numerical scheme of our two-fluid numerical code has to be able to handle large variations in collision frequencies. In the case of dominant collisional terms the equations become stiff and therefore an explicit integration scheme is unstable. 
This is because the collisional terms that appear in the equations of continuity, momentum and energy
  are proportional to the the collision frequency, and the time step
 should be at most the inverse of the collision frequency in order to ensure stability in an explicit scheme. 
This imposes very small time steps when the collisions are dominant. 
A suitable alternative is the use of semi-implicit schemes, similar to that described by \cite{2012Toth}, as the one implemented here.
Our system of equations Eqs. \ref{eq:continuity}, \ref{eq:momentum}, \ref{eq:energy} and \ref{eq:ohm} can be schematically written as follows:
\begin{equation}
\frac{\partial  \vec{U}}{\partial t} = \vec{R}(\vec{U}),
\end{equation}
where $\vec{U}$ is the vector of variables,
\begin{equation} \label{eq:u}
\vec{U}=\{ \rho_n,\rho_c, \rho_n\vec{u}_n,  \rho_c\vec{u}_c,  e_n+\frac{1}{2}\rho_n u_n^2,  e_c+\frac{1}{2}\rho_c u_c^2, \vec{B} \},
\end{equation}
and $\vec{R}(\vec{U})$ are the sum of fluxes and collisional terms. 

\noindent In order to evolve in time this system of equations: we split $\vec{R}$ term into explicit part $\vec{E}$ and implicit part $\vec{P}$, 
\begin{equation}
\vec{R}(\vec{U})=\vec{E}(\vec{U})+\vec{P}(\vec{U})\,.
\end{equation}
The explicit part $\vec{E}(\vec{U})$ contains fluxes, while the implicit part contains the collisional terms, Eqs. (\ref{eq:s}), (\ref{eq:r}) and (\ref{eq:m}). 
With $\vec{U}^*$ being the solution after the explicit part:
\begin{equation}
\vec{U}^*=\vec{U}^n + \Delta t  \vec{E},
\end{equation}
\noindent we solve implicit part using the following discretization:
\begin{equation} \label{eq:beta}
\vec{U}^{n+1} = \vec{U}^{*} + (1-\beta) \Delta t \vec{P}(\vec{U}^{*} ) + \beta \Delta t \vec{P}(\vec{U}^{n+1} ) \,. 
\end{equation}
The value of $\beta$ can vary between 0 and 1, and its influence on the final solution will be discussed below. %

The code solves three systems of two coupled equations in the implicit part, with $\vec{U}$ being in the three cases a vector with two components, as defined above by Eq. \ref{eq:u},  $ $\vec{U}$=\{ \{ \rho_n,\rho_c\}, \{\rho_n\vec{u}_n,  \rho_c\vec{u}_c\},   \{e_n+\frac{1}{2}\rho_n u_n^2,  e_c+\frac{1}{2}\rho_c u_c^2 \} \}$,  so all of the collisional terms, $S_n$, $\vec{R}_n$ and $M_n$
defined in Eqs.  (\ref{eq:s}), (\ref{eq:r}) and (\ref{eq:m})  can be treated in the same way. 
Note however, that, depending on the magnitude of the collisional terms $S_n$ in the continuity equations,  sometimes it is not worth treating them implicitly.  In this case we add these terms into the explicit part $\vec{E}$(\vec{U}).   All the terms on the right hand side of the induction equation, Eq. \ref{eq:ohm}, are treated explicitly.

We linearize $\vec{P}(\vec{U}^{n+1})$ around the value of $U^*$ after an explicit update:
\begin{equation*} 
\vec{P}(\vec{U}^{n+1}) = \vec{P}(\vec{U}^{*} ) +  \hat{J} \cdot \left( \vec{U}^{n+1} - \vec{U}^{*}    \right),
\end{equation*}
where the Jacobian, $\hat{J}$, is defined as follows,
\begin{equation}
\hat{J} = \frac{\partial \vec{P}}{\partial \vec{U}}\,.
\end{equation}

\noindent Due to particular symmetry properties of the equations, $J_{21} = -J_{11} , J_{22} = -J_{12}$.  This allows the solution for the implicit part to be written explicitly.  The update in one sub-step of a two-step Runge-Kutta scheme can then be summarized as:
\begin{equation} \label{eq:toth_comp} 
\vec{U}^{n+\frac{1}{k}} = \vec{U}^{*} + \frac{\Delta t_{\frac{1}{k}} \left ( (1-\beta_{\frac{1}{k}}) \vec{P}^n+ \beta_{\frac{1}{k}} \vec{P}^*  \right )}{1 + \beta_{\frac{1}{k}} \Delta t_{\frac{1}{k}} (\hat{J}_{12} -\hat{J}_{11})}, 
\end{equation}
\noindent where k=2,1; $\Delta_{\frac{1}{k}} = \frac{1}{k} \Delta t$, is the time step used in the substep,  and $\Delta t$ is the complete time step. The scheme parameter $\beta_{\frac{1}{k}}$ can have different value in each substep (subscript $_{\frac{1}{k}}$). The magnitude of $\Delta t$ is limited by the magnetohydrodynamic CFL condition. The explicit part of $\vec{R}$, $\vec{E}$, is  calculated in each Runge-Kutta sub-step using the variables updated after the previous sub-step. The Jacobian $\hat{J}$ is also calculated after an explicit update,  so that $\hat{J}=\hat{J}^*$. 

The elements of the Jacobian can be written in the analytical form.
\noindent For the momentum equations it fulfills that: 
\begin{equation}
\vec{P}^* = \hat{J} \cdot \vec{U}^*.
\end{equation}
For the energy equations the same expression becomes
\begin{equation}
\vec{P}^*=\hat{J} \cdot \vec{U}_T^{*,1} + \hat{K} \cdot \vec{U}_m^{n+\frac{1}{k}},
\end{equation}
where $\vec{U}_m$ and  $\vec{U}_T$ are, correspondingly, the kinetic and internal energy parts of the variable $\vec{U}$. 
The $\hat{J}$ and $\hat{K}$ for the energy equation are defined as follows: 
\begin{eqnarray}
\hat{J} = \frac{\partial \vec{P}}{\partial \vec{U}_T}; \,\,\, \hat{K} = \frac{\partial \vec{P}}{\partial \vec{U}_m}.
\end{eqnarray} 
The variable $\vec{U}_T^{*,1}$ is the difference between $\vec{U}^*$ after an explicit update, and one Runge-Kutta sub-step of the implicit update, i.e.,
\begin{equation}
\vec{U}_T^{*,1} = \vec{U}^* - \vec{U}_m^{n+\frac{1}{k}}.
\end{equation}
 
Overall it means that the energy implicit update is done with variables calculated after the momentum implicit update. The analytical expressions for $\hat{J}$ and $\hat{K}$ are given in the Appendix \ref{app:jac}. The numerical scheme is stable when we use the time step imposed by the CFL. For $\beta_{\frac{1}{2}} = 1$ and $\beta_1 = \frac{1}{2}$  the scheme is formally second order accurate.

\section{Analytical model of waves in a uniform plasma}

Our newly implemented scheme has been checked using simulations of wave propagation in a uniform plasma.  We used two simple analytical cases.  In the first case we consider purely acoustic waves propagating in an atmosphere where  the background equilibrium temperature of charges and neutrals is different.  In this case the difference in behavior between charges and neutrals is created by the different sound speeds  of the background atmosphere. In the second case we consider the propagation of Alfv\'en waves.  Here the difference in the behavior of charges and neutrals is produced additionally due to  the presence of the magnetic field. In this simple setup the problems have analytical  solution that can be obtained by linearizing the two fluid equations and looking for solutions  in the form of  monochromatic waves. The sections below describe the analytical solution and results  for both kinds of simulations, together with the tests of numerical convergence of the scheme.

\subsection{Acoustic waves}

Acoustic waves are longitudinal waves, therefore their velocity vector and direction of propagation are parallel , i.e. $\vec{u}_n \parallel \vec{k}$ and $\vec{u}_c \parallel \vec{k}$.  We choose the direction of propagation along $z$ direction. We neglect the effects of ionization/recombination and thermal exchange, thermal conduction and radiation.  Therefore, we only consider the momentum exchange due to elastic collisions.  In this simple situations the linearized continuity, momentum and adiabatic energy equations become

\begin{eqnarray} \label{eq:tests1}  
\frac{\partial \rho_{\rm n1}}{\partial t} + \rho_{\rm n0} \frac{\partial u_{\rm nz}}{\partial z} &=& 0, \nonumber \\
\frac{\partial \rho_{\rm c1}}{\partial t} + \rho_{\rm c0} \frac{\partial u_{\rm cz}}{\partial z}  &=& 0, \nonumber \\
\rho_{\rm n0} \frac{\partial u_{\rm nz} }{\partial t} + \frac{\partial p_{\rm n1}}{\partial z}& =& \alpha \rho_{\rm n0} \rho_{\rm c0} (u_{\rm cz} - u_{\rm nz} ), \nonumber  \\
\rho_{\rm c0} \frac{\partial u_{\rm cz} }{\partial t} + \frac{\partial p_{\rm c1}}{\partial z} &=&- \alpha \rho_{\rm n0} \rho_{\rm c0}(u_{\rm cz} - u_{\rm nz}),
\nonumber  \\
\frac{\partial p_{\rm n1}}{\partial t} - {c_{\rm n0}}^2 \frac{\partial \rho_{\rm n1}}{\partial t} &=&0, \nonumber \\
\frac{\partial p_{\rm c1}}{\partial t} - {c_{\rm c0}}^2 \frac{\partial \rho_{\rm c1}}{\partial t} &=&0,
\end{eqnarray}
where ${u}_{\rm nz}$ and ${u}_{\rm cz}$ are the velocity projections along $z$ direction; $c_{\rm n0}$  and $c_{\rm c0}$ are sound speeds of neutral and charges. Variables with subscript ``0'' refer to the background unperturbed atmosphere, and variables with subscript ``1'' refer to perturbation. The collisional parameter $\alpha$,  defined by  Eq. (\ref{eq:alpha}) uses the values of the unperturbed mass densities of electrons, ions, neutrals and charges.
Since for the simple model assumed here the value of $\alpha$ only depends on the homogeneous background variables in the linear approximation, it is kept constant in time and space. 

We search for the solutions in the form,
\begin{equation}
\{ u_{\rm nz}, u_{\rm cz}, \rho_{\rm n1},\rho_{\rm c1}, p_{\rm n1}, p_{\rm c1} \} = \{U_n, U_c, R_n, R_c, P_n, P_c\} {\rm e}^{i(\omega t - k z)} ,
\end{equation}
where $\{U_n,U_c, R_n,R_c,P_n,P_c\}$ are,  in general, complex amplitudes for all the perturbed variables. These amplitudes are related through the so-called polarization relation, so one needs to set one of the amplitudes to obtain the rest of them through the following relations:  
\begin{eqnarray} \label{eq:tests_aca}
U_c &=& \frac{\omega R_c}{k \rho_{\rm c0}}; \,\,\, U_n = \frac{\omega R_n}{k \rho_{\rm n0}},  \\
\frac{R_c}{R_n} &=& \frac{i \omega^2 - i k^2 c_{\rm n0}^2 + \alpha \rho_{\rm c0} \omega }{  \alpha \rho_{\rm n0} \omega },  \nonumber \\ 
P_c &=& c_{\rm c0}^2 R_c;  \,\,\, P_n = c_{\rm n0}^2 R_n. \nonumber
\end{eqnarray}
The resulting dispersion relation, which relates the frequency and the wavenumber, has the following form:
\begin{align} \label{eq:tests_acDr}
 i \omega^4 + \alpha \omega^3 (\rho_{\rm c0} + \rho_{\rm n0}) &- i \omega^2 k^2 ( c_{\rm n0}^2 +  c_{\rm c0}^2)  
- \omega k^2 \alpha (\rho_{\rm c0}  c_{\rm c0}^2 \nonumber \\&  + \rho_{\rm n0}  c_{\rm n0}^2  ) 
 + i k^4 c_{\rm n0}^2 c_{\rm c0}^2  = 0. 
\end{align}
Once the background variables are fixed, this also fixes the value of the collisional parameter $\alpha$. It is then convenient to operate in terms of the following parameters: $\omega/k$, and $\alpha/k$. The latter ratio is an effective measure of the collisional strength. The analysis below will be done in terms of the adimensional variables:
\begin{eqnarray} \label{eq:adim_ac}
E = \frac{\omega}{k c_{\rm tot}} ;  \,\,\, 
F = \frac{\alpha \rho_{\rm tot}}{k c_{\rm tot}}.
\end{eqnarray}
These variables are similar to those used in \cite{2013Soler}. Later on in the paper we represent the solution of the dispersion relation in terms of these adimensional variables, keeping in mind that by ``varying the collisional strength $F$'' we understand varying the wavenumber $k$, since the value of $\alpha$ is fixed. 
 
Using adimensional variables, the polarization relations, and the dispersion relation become:
\begin{equation} 
 \frac{U_c}{R_c} = E \frac{c_{\rm tot}}{\rho_{\rm c0}} ;  \,\,\,\,\, 
\frac{R_n}{R_c} = \frac{E F \rho_{\rm n0}/\rho_{\rm tot} }{i E^2 - i {c_n}_0^2/c_{\rm tot}^2 + E F \rho_{\rm c0}/\rho_{\rm tot}}, 
\end{equation}
\begin{equation} \label{eq:adim_disp}
 i E^4 + F E^3  - i E^2 a_2  - E F    + i a_0   = 0,
\end{equation}
\noindent where:
\begin{equation} 
a_2 = \frac{ c_{\rm n0}^2 +  c_{\rm c0}^2}{c_{\rm tot}^2} ;  \,\,\,\,\,
a_0 = \frac{c_{\rm n0}^2 c_{\rm c0}^2}{c_{\rm tot}^4}.
\end{equation}
Here $\rho_{\rm tot} =  \rho_{\rm n0} + \rho_{\rm c0} $ and $c_{\rm tot} = \sqrt{\gamma (p_{n_{0}} + p_{c_{0}})/(\rho_{\rm n0} + \rho_{\rm c0})}$ are the total density and the sound speed of the whole fluid. 
%
\begin{figure*}[t]
\includegraphics[width = 8.5cm]{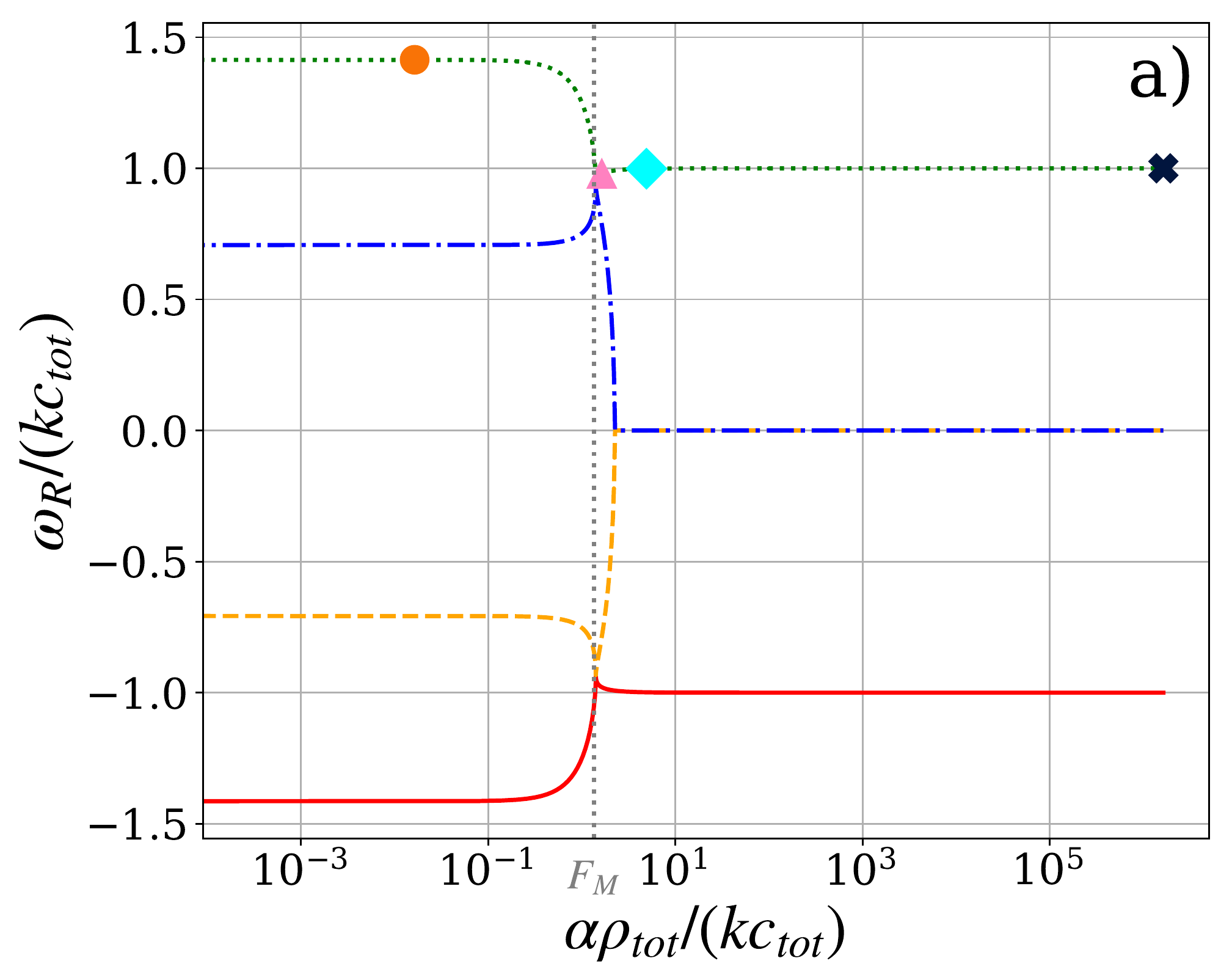}
\includegraphics[width = 8.5cm]{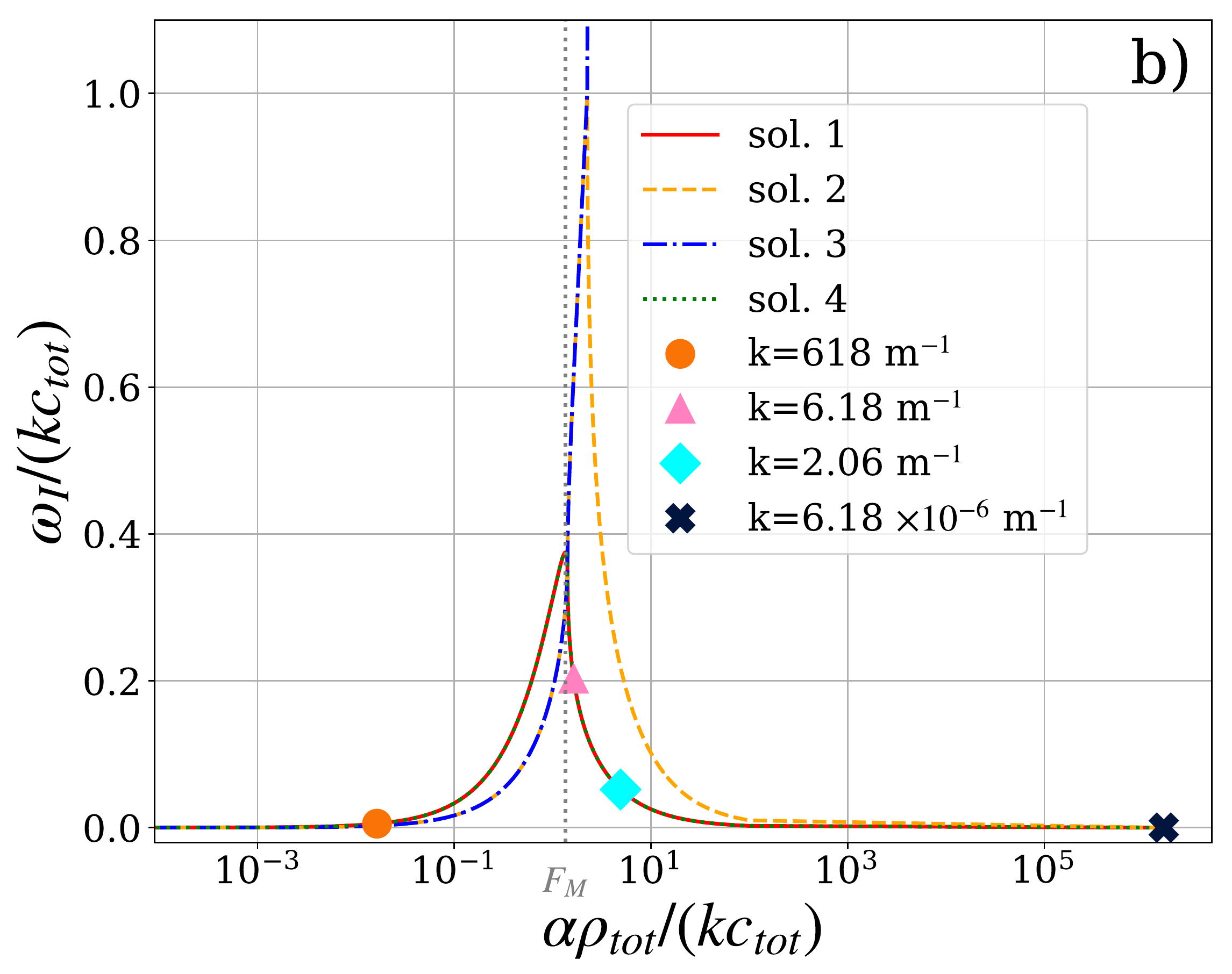}
\caption{ Solutions of the dispersion relation, Eq. \ref{eq:tests_acDr}, for the acoustic wave case.  
Left (a) and right (b) panels  present the real and imaginary parts of the dimesionless wave frequency $\omega$, correspondingly,  as a function of the dimensionless wavenumber  k. The axis are scaled in non-dimensional units E and F, defined in Eq. (\ref{eq:adim_ac}) for better visualization. Four solutions are marked with different colors and line styles: solid, red line for solution 1, orange, dashed line for solution 2, blue, dot-dashed line for solution 3, and green, dotted line for solution 4.  Note that solutions 1 and 4 have the same imaginary parts and are superposed at the right panel.  Solution 2 and 3 have the same real part equal to zero for large $\alpha$/k and also superpose.  
The values of k for which we compare the analytical and numerical solution, using the values of $\omega$, 
corresponding to the solution 4 (green dots), are marked in the panels:
orange circle (k=618 m$^{-1}$), pink triangle up (k=6.18 m$^{-1}$), cyan diamond (k=2.06 m$^{-1}$), and
black "X" (k=6.18 $\times$ 10$^{-6}$ m$^{-1}$).
} 
\label{fig:acousticw}
\end{figure*}
\subsection{Alfv\'en  waves}
%
Alfv\'en waves are transversal and incompresible waves, polarized perpendicularly to the direction of the background magnetic field.  The analytical solution for the Alfv\'en wave propagation in partially ionized plasma used in our work is similar to those  developed previously by \cite{2013Soler}.  We consider here a simple case of Alfv\'en waves with propagation only along the background magnetic field.  These are frequently called pure Alfv\'en waves.  In our case, $\vec{k} \parallel \vec{B_0}, \vec{u}_n \parallel \vec{B_1},  \vec{u}_c \parallel \vec{B_1},  \vec{B_1} \perp  \vec{B_0}$.  We choose the equilibrium magnetic field along $z$ direction and perturbation of magnetic field and velocities along $x$ direction. 
\begin{eqnarray*}
\vec{B}_0&=&(0,0,B_0), \,\,\, \vec{k}=(0,0,k), \,\,\,  \vec{B}_1=(B_{\rm 1x},0,0), \\
\vec{u}_{c} &=& ({u_{\rm cx}},0,0), \,\,\,  \vec{u}_{n}=({u_{\rm nx}},0,0)
\end{eqnarray*}
Otherwise the assumptions of the equations are similar to the acoustic wave case. The linearized momentum and induction equations become as follows,
\begin{eqnarray} \label{eq:tests22}
\rho_{\rm n0} \frac{\partial u_{\rm nx} }{\partial t} &=&  \alpha \rho_{\rm n0} \rho_{\rm n0} (u_{\rm cx} - u_{\rm nx} ), \nonumber \\
\rho_{\rm c0} \frac{\partial u_{\rm cx} }{\partial t} &=& \frac{B_0}{\mu_0}\frac{\partial B_{\rm 1x}}{\partial z}
- \alpha \rho_{\rm c0} \rho_{\rm n0} (u_{\rm cx} - u_{\rm nx} ), \nonumber \\
\frac{\partial B_{\rm 1x}}{\partial t} &=& B_0\frac{\partial u_{\rm cx}}{\partial z}.  
\end{eqnarray}
\noindent The equations of continuity and energy conservations are not needed in this case since the Alfv\'en waves are purely incompressible. Similar to acoustic waves we use the solutions of the form:
\begin{equation}
\{ u_{\rm nx}, u_{\rm cx}, B_{\rm 1x}\} = \{U_n,U_c, B\} {\rm e}^{i(\omega t - k z)},
\end{equation}
where the amplitudes can be  complex in general. Plugging this  ansatz into Eqs. \ref{eq:tests22}, we obtain the following polarization relations:
\begin{eqnarray} \label{eq:alf_an1}
B  &=& - \frac{k {U_c}{B_0}}{\omega}, \nonumber \\
\frac{{U_c}}{{U_n}} &=& \frac{i \rho_{\rm n0}  \omega + \alpha \rho_{\rm n0}  \rho_{\rm c0} }{ \alpha \rho_{\rm n0}  \rho_{\rm c0}} \,.
\end{eqnarray}
\noindent and the dispersion relation in the following form:
\begin{equation} \label{eq:tests_alfDr}
-\mu_0 \rho_{\rm c0} \omega^3 + i \mu_0 \alpha   \rho_{\rm c0}  ( \rho_{\rm c0} +  \rho_{\rm n0}) \omega^2 +
{B_0}^2  k^2 \omega -i  \alpha \rho_{\rm c0} {B_0}^2  k^2 = 0 .
\end{equation}
For the same reason explained for the acoustic waves, we will vary the collision strength by varying
the wavenumber k parameter  to study its influence on the propagation properties of the waves.
\noindent Using the adimensional variables:
\begin{eqnarray} \label{eq:adim_alf}
C = \frac{\omega}{k {v_A}_{0}}; \,\,\,
D = \frac{\alpha \rho_{\rm tot}}{k {v_A}_{0}},
\end{eqnarray}
\noindent the dispersion relation, and the polarization relations are:
\begin{equation} \label{eq:disp_alf_adim}
-C^3 + i D C^2 + C - i D \frac{\rho_{\rm c0}}{\rho_{\rm tot}} = 0,
\end{equation}
\begin{equation}
\frac{B}{U_c} = -\frac{B_0}{C {v_A}_0}, \,\,\,
\frac{U_c}{U_n} = 1 + i \frac{C}{D} \frac{\rho_{\rm tot}}{\rho_{\rm c0}}.
\end{equation}
%
\begin{figure*}
\includegraphics[width = 8.5cm]{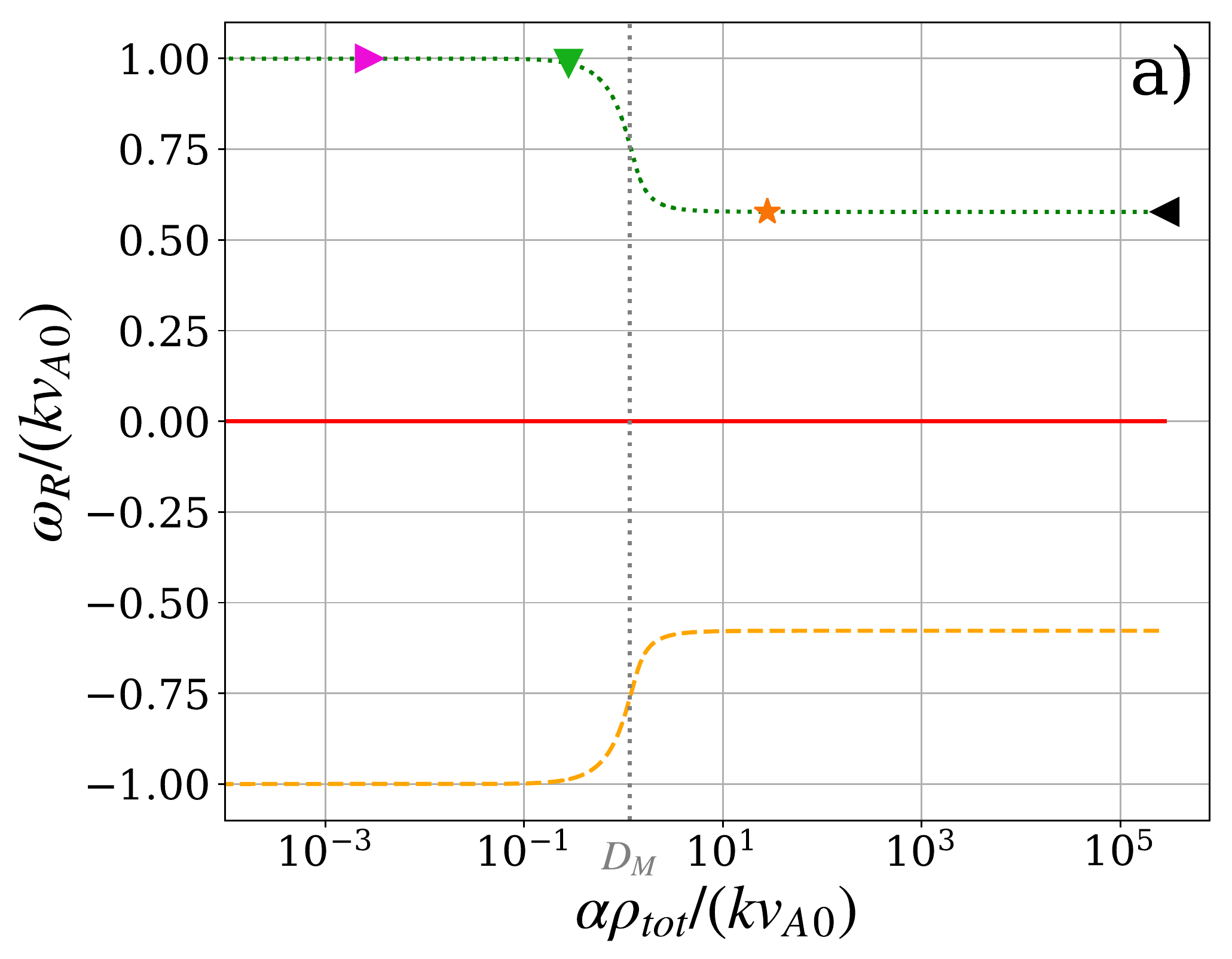}
\includegraphics[width = 8.5cm]{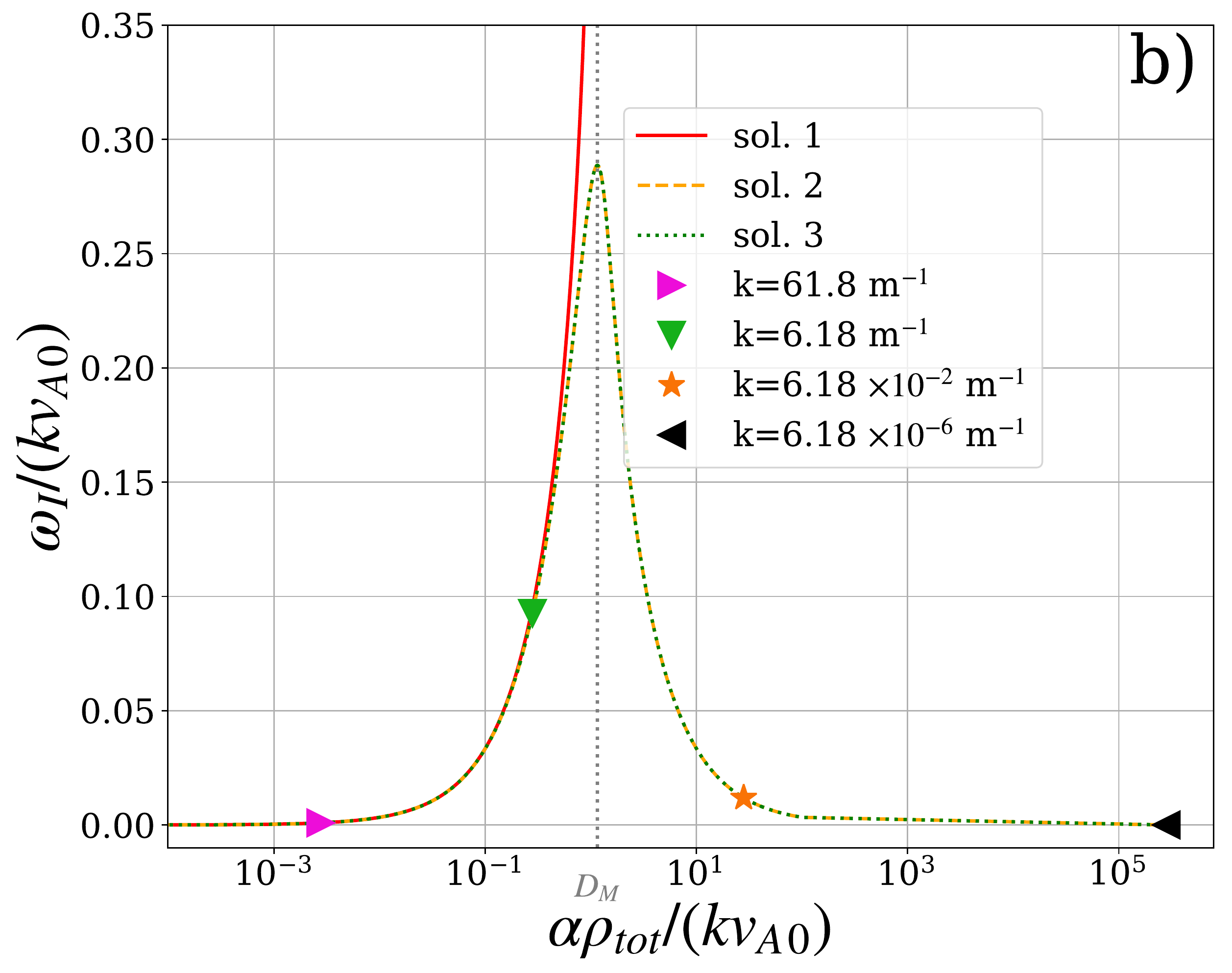}
\caption{Solutions of the dispersion relation, Eq. \ref{eq:tests_alfDr}, for the Alfv\'en wave case.  Left (a) and right (b) panels  present the real and imaginary parts of the wave  frequency $\omega$, correspondingly,  as a function of the wavenumber k. The axis are scaled in the non-dimensional units: C and D, defined in Eq. (\ref{eq:adim_alf}) for better visualization. Three solutions are marked with different colors and line styles: red, solid line for solution 1, orange, dashed line for solution 2, and green, dotted line for solution 3. 
The values of k for which we compare the analytical and numerical solution, using the values of $\omega$, 
corresponding to the solution 3 (green dots) are marked in the panels:
pink triangle right (k=61.8 m$^{-1}$), green triangle down (k=6.18 m$^{-1}$), orange star (k=6.18 $\times$ 10$^{-2}$ m$^{-1}$), and
black triangle left (k=6.18 $\times$ 10$^{-6}$ m$^{-1}$).
} 
\label{fig:alfvenw}
\end{figure*}
%
\subsection{Background atmosphere}  \label{subsec:eq}
%
The values of the uniform background atmosphere used in this study are given in Table \ref{tab:bg_tests}.  
For the Alfv\'en wave tests, a z-directed magnetic field $B_0 = 5 \cdot 10^{-3} T$  is also added. 
The corresponding Alfv\'en speed, calculated using the density of charges is, ${v_A}_0 = {B_0}/{\sqrt{\mu_0 \rho_{\rm c0}}}=6.32 \cdot 10^4$ m/s. 
The collisional parameter $\alpha$ defined in Equation (\ref{eq:alpha}) calculated for  the background has the value  $7.324 \times 10^{12}$ m$^3$/kg/s. 
The electrons are not taken into account ($n_{\rm c0}=n_{\rm i0}$), 
they do not contribute to the pressure of charges, and the effective neutral-charge and charge-neutral collisional frecuencies
defined in Equation (\ref{eq:eff_coll}) become the neutral-ion and ion-neutral collision frequency, with values $\nu_{ni} \approx 3.65 \times 10^4$ s$^{-1}$,  
and $\nu_{in} \approx 7.3 \times 10^4$ s$^{-1}$, respectivelly.
%
\begin{table}
\caption{Values of the background atmosphere parameters used in the tests of the wave propagation in the uniform medium for neutrals (left) and charges (right).}
\begin{tabular}{ll}
\hline			
Neutrals & Charges \\
\hline			
$n_{\rm n0}=6 \cdot 10^{18} m^{-3}$  & $n_{\rm c0}= \frac{1}{2}n_{\rm n0} =  3 \cdot 10^{18}  m^{-3}$   \\
$\rho_{\rm n0} = 9.96\cdot 10^{-9} kg/m^3$ &  $\rho_{\rm c0} = \frac{1}{2} \rho_{\rm n0}= 4.98 \cdot 10^{-9} kg/m^{3}$ \\
${p_n}_{0} = 0.35$ Pa &  ${p_c}_{0} = 2 {p_n}_{0}  =$ 0.7 Pa\\
$c_{\rm n0} = 7.65 \cdot 10^3$ m/s & $c_{\rm c0}$ ={2 ${c_n}_0$} = 1.53 $\cdot 10^4$ m/s \\
${T_n}_0 = 4227$ K & ${T_c}_0 = 4 {T_n}_0 = 16908$ K\\
\hline  
\end{tabular} \label{tab:bg_tests}
\end{table}
%
\subsection{Parameters of the perturbation} \label{subsec:pert}

We have to supply several free parameters to calculate the solution. We have chosen the real wave number $k$, and obtain the complex $\omega$ from the dispersion relations, Eqs. (\ref{eq:tests_acDr}) and (\ref{eq:tests_alfDr}). This means that we consider wave damping in time. Additionally, since the perturbations in all the magnitudes are related, we need to supply the amplitude of one of them in order to calculate the others. Generally, since the polarization relations are complex, there will be the phase shift between oscillations of different quantities.

We have selected  the wave number to be $k = {2\pi n}/{L_z}$ m$^{-1}$, with $n$ being an integer number to adjust the fixed number wavelengths to the size of numerical domain $L_z$. We vary the wavenumber k by varying the domain length $L_z$ in order to have n = 11 (a random choice) for all the simulations.
 
For the acoustic waves we will choose the amplitude of the perturbation in density of charges: $R_c = 10^{-3} \rho_{\rm c0}$,  and for the Alfv\'en waves we choose  the amplitude of the perturbation in velocity of charges: $U_c=10^{-5} v_{A_0}$.  We calculate the other amplitudes from  Eqs. (\ref{eq:tests_aca}) for the acoustic waves and Eqs. (\ref{eq:tests22}) for  the Alfv\'en waves.  In all the cases, for the numerical solution we have used periodic boundary conditions.  For the initial condition of the perturbation we used the  analytical solution at t = 0 s in the whole domain.

We run all simulations for a total time $t_F$ so that we have $L_z/t_F$ equal to 7447.273 m/s for all the simulations of  acoustic waves and equal to 32125.49 m/s for the  Alfv\'en waves simulations. These values are of order of the characteristic speeds: the sound speed and the Alfv\'en speed.

\subsection{Solution of the dispersion relation for acoustic waves}
%
The acoustic wave dispersion relation is a fourth order equation in both  $\omega$ and k.  
If the neutrals and charges have the same background temperature, the dispersion relation has 
four solutions, with simple expressions, as shown in Equation (\ref{eq:solAWsameT}).
In our case $T_{c0}/T_{n0} = 4$, the values of the coefficients
in Equation (\ref{eq:adim_disp}) are: $a_2$ = 5/2 and $a_0$ = 1, and the expressions for the solutions, 
obtained using Mathematica software, in this case, are:
\begin{align} \label{eq:solMaw}
E_{4,1} &=& \frac{1}{4}\left(i F - \sqrt{2-F^2} \pm \sqrt{
18 - 2 F^2 - 2 i F \sqrt{2 - F^2}}  \right ), \nonumber \\
E_{3,2} &=& \frac{1}{4}\left(i F + \sqrt{2-F^2} \pm \sqrt{
18 - 2 F^2 + 2 i F \sqrt{2 - F^2}}  \right ). 
\end{align} 
We compute the  complex  frequency, $\omega$,  for the values of k between $6180$ and $6.18 \times 10^{-6}$ m$^{-1}$, for each of the four 
expressions of the mathematical solutions shown in Equation (\ref{eq:solMaw}), and marked as sol. 1, 2, 3, 4 in Figure \ref{fig:acousticw}.  
The real and imaginary parts of the four solutions are shown in the normalized units, E = f(F),  defined in Equation (\ref{eq:adim_ac}), 
i.e. $\omega_{\{I,R\}}/k/ {c_{\rm tot}}$  as a function of $\alpha {\rho_{\rm tot}}/k/{c_{\rm tot}}$,  for better visualization. 
The solutions with positive $\omega_R$ travel in the positive direction of the $z$ axis, and those with negative $\omega_R$ travel in the negative direction.
 
Waves corresponding to the solutions 1 and 4  (red and green colors) propagate with a speed ($v_{ph} = {\omega_R}/{k}$) equal to the sound speed of the charges ($c_{\rm c0}/c_{\rm tot} = \sqrt{2} \approx1.414$) for small ratio $\alpha/k$, in the negative and positive direction of the $z$ axis, correspondingly.  For the large values of $\alpha$/k their propagation speed becomes that of the whole fluid, $c_{\rm tot}$.  
The imaginary part of the frequency, $\omega_I$ is positive meaning  wave damping. The imaginary part would be zero if the charges and neutrals had the same background temperature, since in that case the solutions of the dispersion relation, Eq. \ref{eq:adim_disp}, would simplify to:
\begin{eqnarray} 
E_{4,1} &=& \pm 1,  \label{eq:solAWsameT} \nonumber \\
\end{eqnarray}
with zero imaginary part for the solutions 1 and 4. Since the only difference between neutrals and charges in this particular model is in their sound speed, it is not surprising that making the sound speed the same the damping disappears since both fluids would oscillate with exactly the same velocity and therefore the frictional damping term would become strictly zero.

The value of damping is the same for solutions 1 and 4. For either very low or very high values of the ratio $\alpha$/k, the damping  relative to the wavenumber ($\omega_I/k$) is small and approaching zero.  The value of the ratio $\omega_I/k$ is maximum  at a point  located on the x axis at $F_M =\alpha {\rho_{\rm tot}}/k_M/{c_{\rm tot}} = \sqrt{2}$,  and corresponds to $k_M \approx 7.5$ m$^{-1}$. E($F=F_M$) = $1/4(\sqrt{14} + i \sqrt{2})$, this gives the value of  $\omega_R(F=F_M)/(\alpha \rho_{\rm n0}) = 3\sqrt{7}/8 \approx 1$, which means that the real part of the wave frequency is approximately equal to the ion-neutral collision frequency at the point where the damping relative to wavenumber, the ratio $\omega_I/k$, is maximum ($F_M$). 

Waves corresponding to solutions 2 and 3 (yellow and blue colors) propagate with a speed equal  to the sound speed of neutrals ($c_{\rm n0}/c_{\rm tot} = \sqrt{2}/2 \approx 0.707$) for the small values of $\alpha$/k. For large $\alpha$/k, these solutions have zero propagation speed (zero $\omega_R$). The ratio, $\omega_I/k$, is again small for  weak or strong  collisional coupling (small or large $\alpha$/k), and  has the maximum located on the x axis at a point very close to $F_M$.

These results are easy to interpret. For small collisional frequencies compared to the real part of  the wave  frequency, i.e. weak collisional coupling, the propagation of the waves mostly depends only on the properties of either of the fluids. Since the neutrals and the charges have different sound speeds, in the case of weak collisional coupling the waves propagate either at the sound speed of neutrals or of charges. Correspondingly, if one perturbs charges with a certain amplitude, the weak drag forces will translate some of this perturbation to neutrals, but the amplitude of this perturbation will be very small, as indeed follows from the polarization relations, Eqs. \ref{eq:tests_aca}. The opposite is also true. 

For large collision frequencies compared to the real part of the wave frequency, i.e. strong collisional coupling, the charges and the neutrals become coupled and the wave propagates at the sound speed of the whole fluid. Then, the amplitudes of the velocities of neutrals and charges are equal.

For the numerical tests we will only show the results corresponding to the solution 4  (the green lines in Figure  \ref{fig:acousticw}), i.e. the solution for waves which propagate at the sound speed of the  charges for small $\alpha$/k. The results in other cases are similar. We have selected one wavenumber so that $\alpha {\rho_{\rm tot}}/k/{c_{\rm tot}}$ is less than $F_M$, and three values of k, for which this quantity is larger than $F_M$. These frequencies are marked  with symbols in Figure \ref{fig:acousticw}. In order to relate the results to observations, Table \ref{tab:charValAW} provides the period, P=$2 \pi/\omega_R$, and the damping time, $T_D = 2 \pi/\omega_I$ for the waves used in the simulations (the wavenumbers k  marked in Figure \ref{fig:acousticw}),  and for the wave corresponding to the maximum damping relative  to the wavenumber (with wavenumber $k_M$) in physical units.  We can see that the waves have extremely short periods and wavelengths and do not correspond to the  waves observed in the chromosphere, which have typical frequencies of 3-5 mHz. This is due, primarily to our use of a uniform, unstratified , background atmosphere which does not correspond to the chromosphere. The temperatures of charges and neutrals are different by a factor of 4, and this situation is also unrealistic. For these reasons, the value of the collisional parameter $\alpha$, corresponding to the background atmosphere,  may be unphysical for the chromosphere. Moreover, for the purposes of code verification we wanted to test both uncoupled and strongly  coupled cases, and the values of k that we have chosen cover the full range, even if the waves corresponding to the largest values of k have  small temporal and spatial scales that cannot be observed.
%
\begin{figure*}
\centering
\includegraphics[width = 8.5cm]{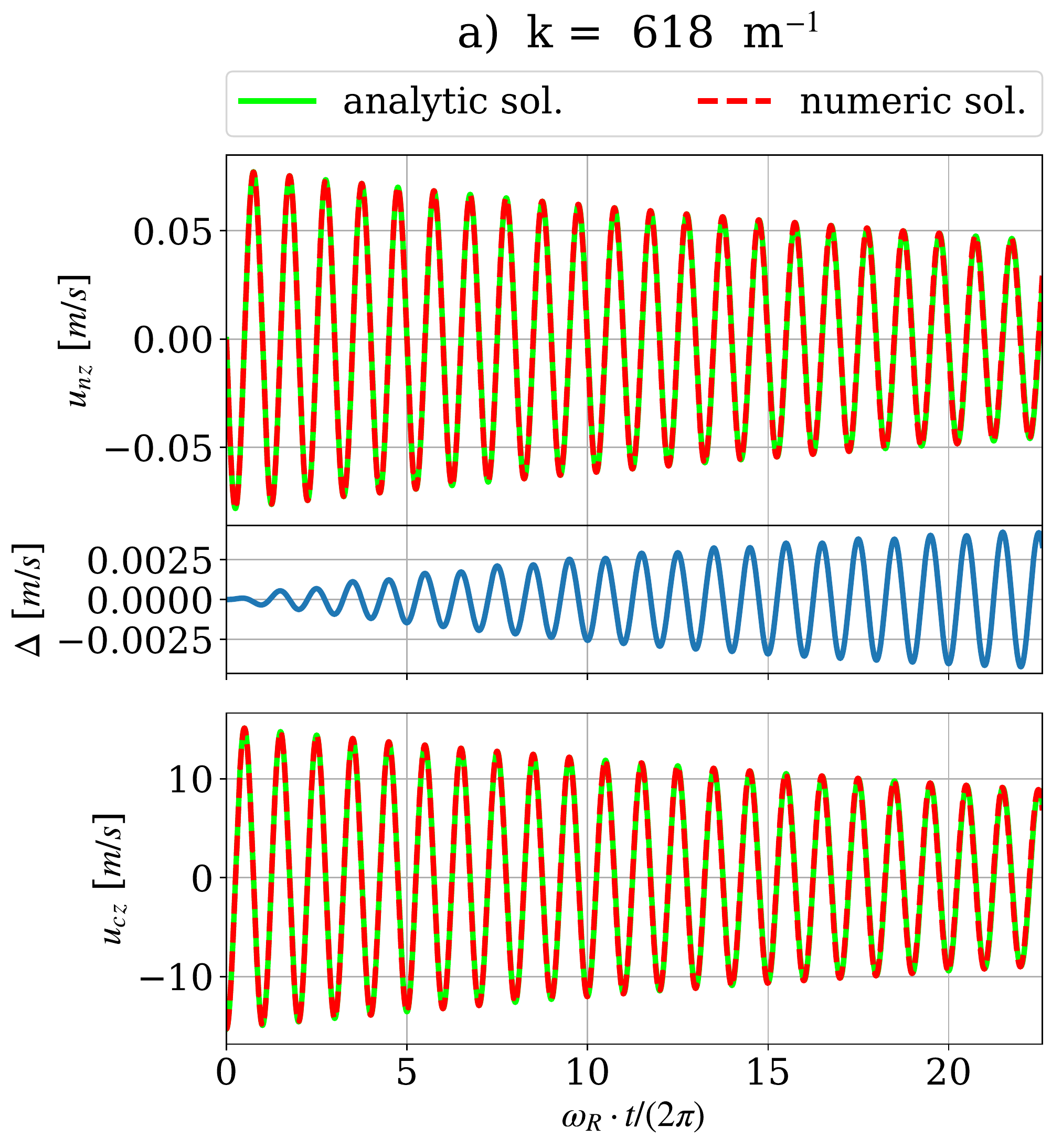}
\includegraphics[width = 8.5cm]{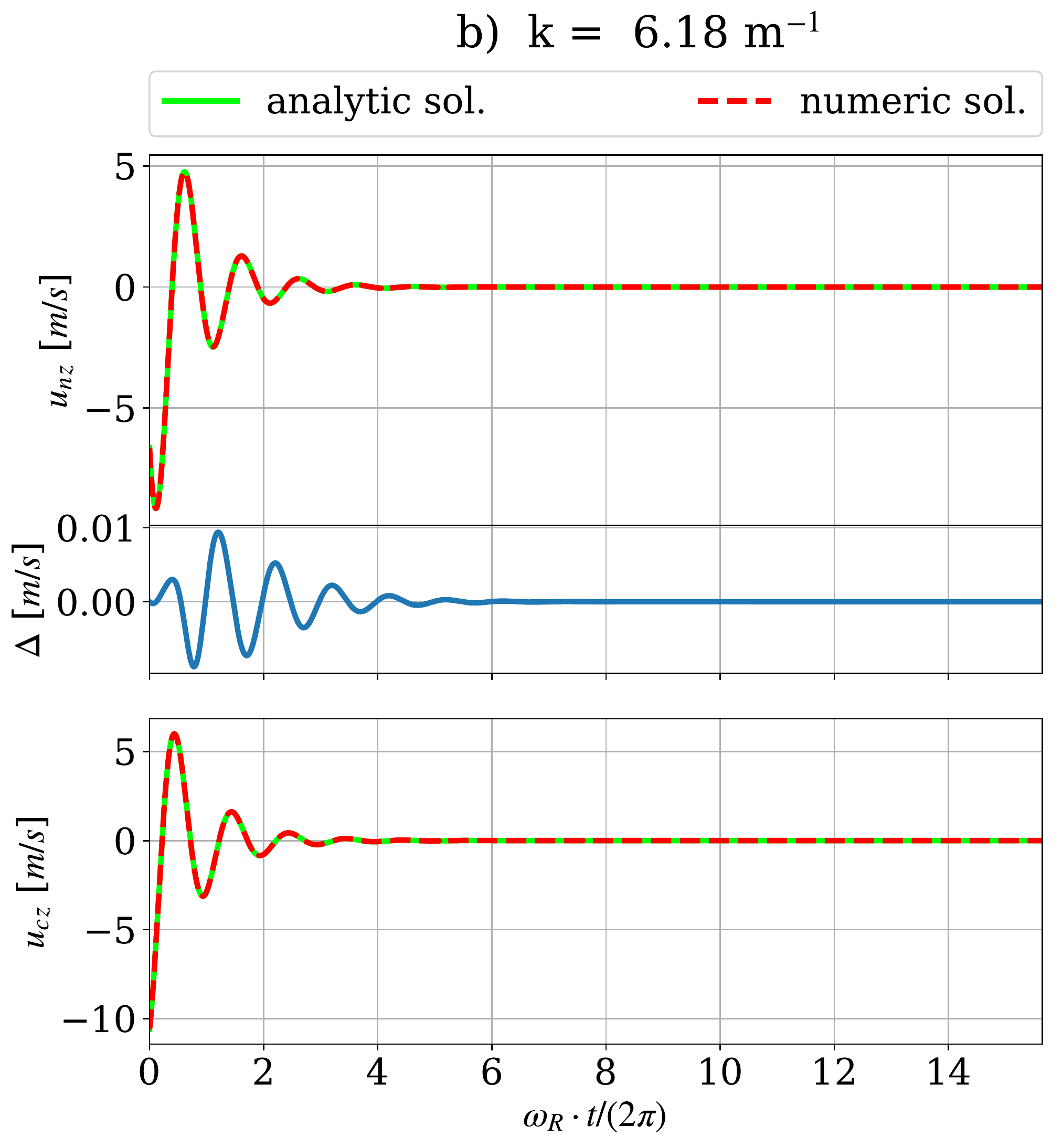}

\includegraphics[width = 8.5cm]{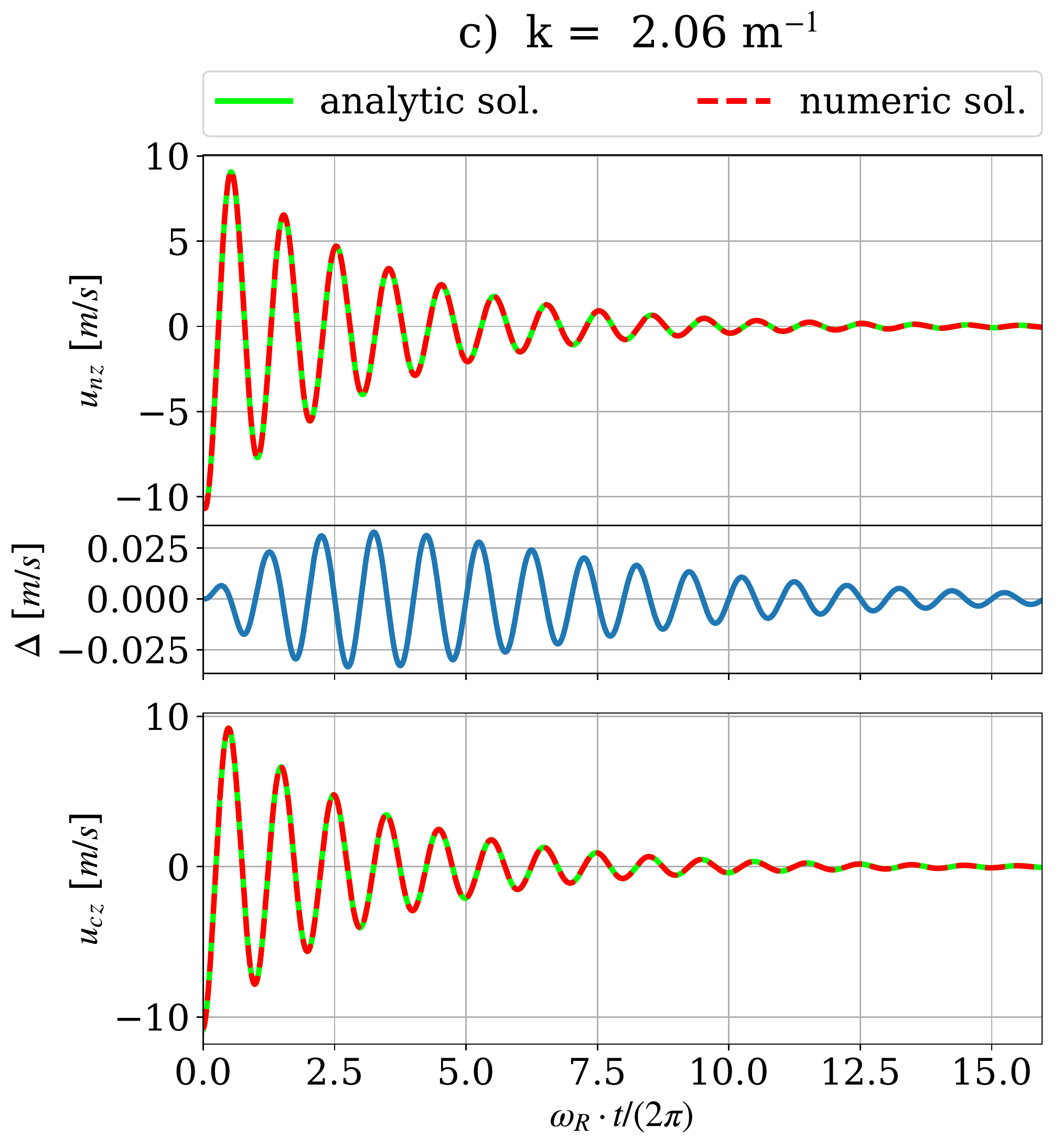}
\includegraphics[width = 8.5cm]{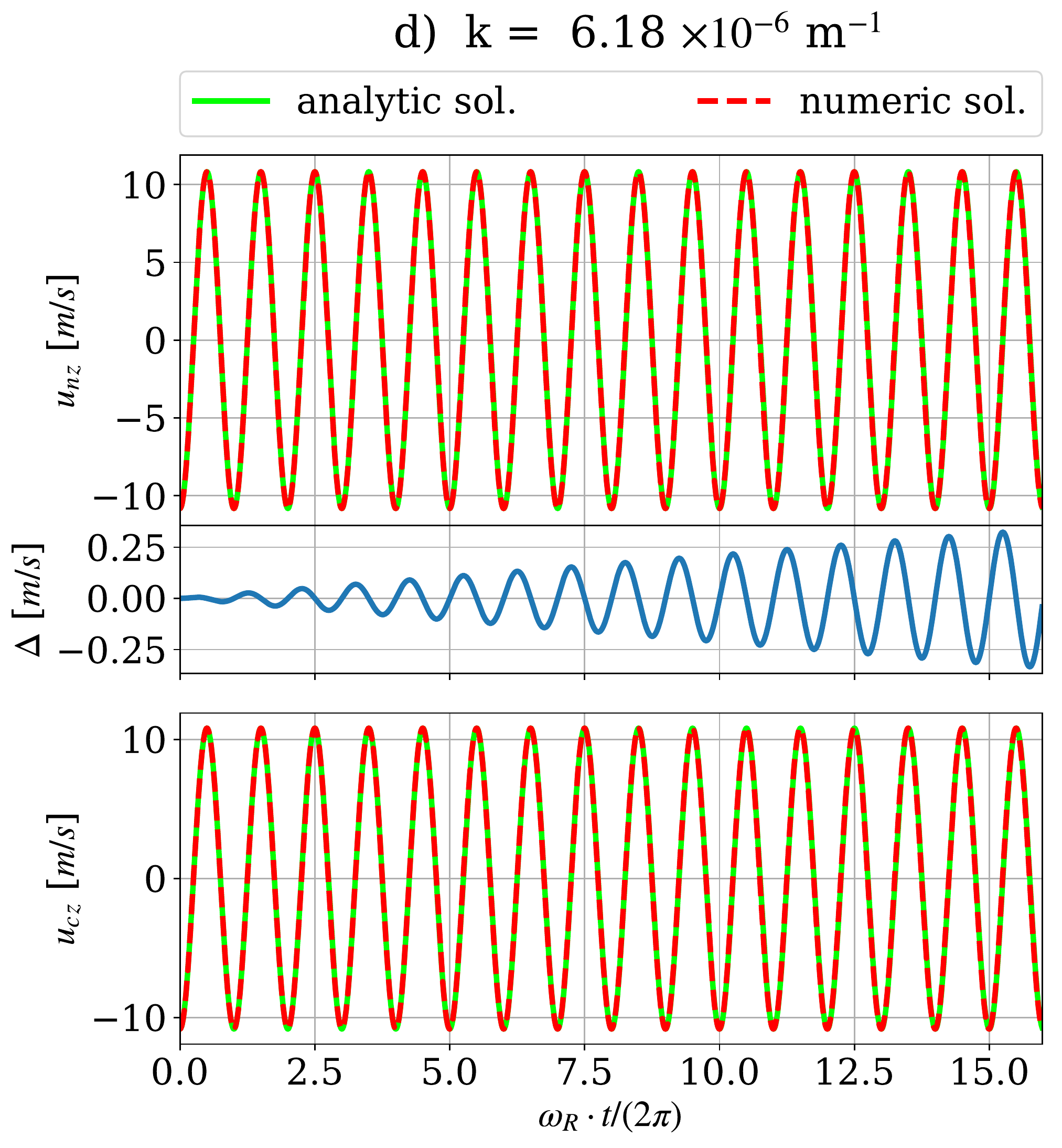}
\caption{Simulations of acoustic waves propagation in a homogeneous plasma. Numerically (red, dashed line) and analytically (green, solid line) calculated time evolution of the velocity of neutrals 
and charges as a function of time. 
 Below the panels showing the velocity of neutrals, the difference ($\Delta$) between the numerical solution and the analytical solution for $u_{\rm nz}$ is given.
Time is measured in units of the wave period, $2\pi/\omega_R$. Panels from left to right and from top to bottom show simulations for different values of the wavenumber k, as indicated in the legend of Fig. \ref{fig:acousticw}. }
\label{fig:figAcT}
\end{figure*}
%
\subsection{Solution of the dispersion relation for Alfv\'en waves}
%
\noindent As in the case of the acoustic waves, we solve the dispersion relation for values of k between $6180$ and $6.18 \times 10^{-6}$ m$^{-1}$, and we plot the real and imaginary part of the wave frequency in Figure \ref{fig:alfvenw}, as a function of the wavenumber  in non-dimensional units, C = f(D), defined in Equation (\ref{eq:adim_alf}), i.e. $\omega_{\{I,R\}}/k/{v_A}_0$  as function of $\alpha \rho_{\rm tot} /k/{v_A}_0$. The dispersion relation is a third order equation in $\omega$ and there are three solutions.
In our case, $\rho_{\rm c0}/\rho_{\rm tot} = 1/3$, and the solutions obtained using Mathematica software, are:
\begin{eqnarray} \label{eq:solMalf}
C_1 &=& \frac{i}{3} \left (D + \frac{D^2-3}{G} + G \right ), \nonumber \\
C_{3,2} &=& \frac{i}{3} \left (D +  L_{3,2}^0 i^{-4/3} \frac{D^2-3}{G} - L_{3,2}^1 i^{-2/3} G  \right ).	
\end{eqnarray}
with:
\begin{eqnarray}
G &=& \left (D^3 + 3 \sqrt{3 - 3 D^2 + D^4} \right)^{1/3}, \nonumber  \\
L_{3}^0 &=& -i \sqrt{3}/2 - 1/2;\,\,\,L_{2}^0 = 1,  \nonumber \\
L_{3}^1 &=& i \sqrt{3}/2 - 1/2;\,\,\,L_{2}^1 = 1 \,.
\end{eqnarray}
They represent  one wave which does not propagate ($\omega_R = 0$, red color), one wave traveling in the positive direction of $z$ axis (green color)  and a similar wave traveling in the negative direction (yellow color).

For small values of the collisional frequency compared to the real part of the wave frequency the propagation speed ($v_{ph} = {\omega_R}/{k}$) is the Alfv\'en speed considering only the density of charges (${v_A}_0$). For the large values of $\alpha$/k, the value of $v_{ph}$ is the Alfv\'en speed of the whole plasma considering both neutrals and charges, ${v_A}_{tot}$. Note that $ {v_A}_{tot}/ {v_A}_0 = \sqrt{{\rho_{\rm c0}}/{\rho_{\rm tot}}} = 1/\sqrt{3} \approx 0.577$.  With no collisions between charges and neutrals, the imaginary part of solutions 2 and 3 vanishes, 
\begin{equation} \label{eq:solAlfnoN}
C_{3,2} = \pm 1,
\end{equation}
meaning that the damping is related to the presence of neutrals. In our case, the imaginary part  is positive,  and the maximum damping relative to the wavenumber, the ratio $\omega_I/k$,  is located at a point $D_M = \alpha \rho_{\rm tot} /k_M^A/v_{A0} = 2/\sqrt{3}$ on the x axis, correspondig to a value of $k_M^A \approx 8.8$ m$^{-1}$. This gives the value of $\omega_R(D=D_M)/(\alpha \rho_{\rm n0})  \approx 1$.  As in the case of the acoustic waves, the real part of the wave frequency is almost equal to the ion-neutral collision  frequency at the point where the damping relative to wavenumber is maximum. Otherwise, the damping relative to the wavenumber  is very small. 

For the numerical tests below we will use the solution traveling in the positive direction of the $z$ axis (green color in Fig. \ref{fig:alfvenw}) and for values of the wavenumbers k indicated in the legend of Fig.  \ref{fig:alfvenw}.
The period and the damping time for the waves used in the simulations (corresponding to the values of 
k marked  in Figure \ref{fig:alfvenw}), and for the wave corresponding to the maximum damping relative 
to the wavenumber (with wavenumber $k_M^A$) are given in Table \ref{tab:charValAlf}.
For the same reasons as  the acoustic wave tests, the frequencies and the  wavenumbers of the Alfv\'en waves do not correspond to typically observed waves in the chromosphere. 
%
\begin{table} [b]
\begin{center}
\caption{Values of the period, and damping time for the waves used in the simulations
and the wave with the maximum damping relative to the wavenumber for the case of the acoustic waves.}
\begin{tabular}{c c c c }
\multicolumn{3}{l}{\bfseries Acoustic waves}\\
\hline
k [m$^{-1}$]  & $P=2 \pi/\omega_R$ [s]   &  $T_D=2 \pi/\omega_I$ [s]\\
\hline      
618  & $6.65 \times 10^{-7}$ & $1.72 \times 10^{-4}$ \\
7.5 ($k_M$)  & $7.36 \times 10^{-5}$ & $2.07 \times 10^{-4}$ \\
6.18  & $9.6 \times 10^{-5}$ & $4.63 \times 10^{-4}$ \\
2.06  & $2.82 \times 10^{-4}$ & $5.43 \times 10^{-4}$ \\
$6.18\times 10^{-6}$ &   $94.5$   &  $6.15 \times 10^8$ \\
\hline  
\end{tabular}
\label{tab:charValAW}
\end{center}
\end{table}
\begin{table} [b]
\begin{center}
\caption{Values of the  period, and damping time for the waves used in the simulations 
and the wave with the maximum damping relative to the wavenumber for the case of the Alfv\'en waves.}
\begin{tabular}{c c c c }
\multicolumn{3}{l}{\bfseries Alfv\'en waves}\\  
\hline    
k [m$^{-1}$]  & $P=2 \pi/\omega_R$ [s]  &  $T_D=2 \pi/\omega_I$ [s]\\
\hline      
61.8 & $1.6 \times 10^{-7}$ & $1.72 \times 10^{-4}$ \\
6.18  & $1.63 \times 10^{-5}$ & $1.73 \times 10^{-4}$ \\
8.8 ($k_M^A$)& $8.67 \times 10^{-5}$ & $2.3 \times 10^{-4}$ \\
$6.18\times 10^{-2}$  & $2.79 \times 10^{-3}$ & 0.135 \\
$6.18\times 10^{-6}$ & 27.89   &  $1.35 \times 10^{7}$ \\
\hline  
\end{tabular}
\label{tab:charValAlf}
\end{center}
\end{table}
%
\section{Results of numerical calculations}
%
We have run simulations of acoustic and Alfv\'en waves  using as initial conditions the equilibrium atmosphere and perturbation
described in the previous sections: \ref{subsec:eq} and \ref{subsec:pert}.
We have used  several values of the wavenumber k marked in the corresponding panels described in the above section. In order to observe the damping in time we choose a point located at 1/2 of the domain and plot the evolution of  the variables at the same point as function of time. The numerical solution is then compared to the analytical solution.
%
\begin{figure*}
\centering 
\includegraphics[width = 8.5cm]{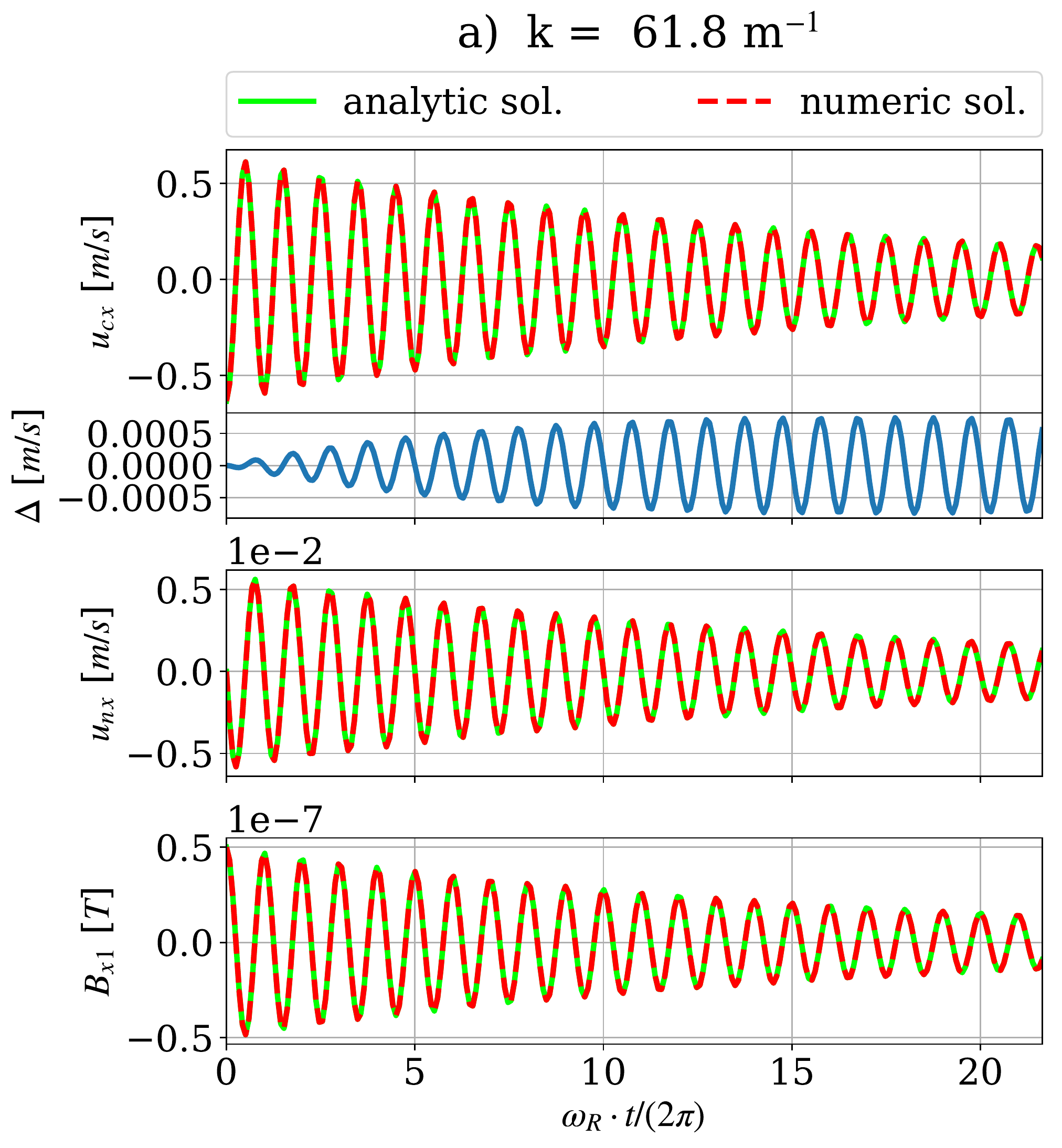}
\includegraphics[width = 8.5cm]{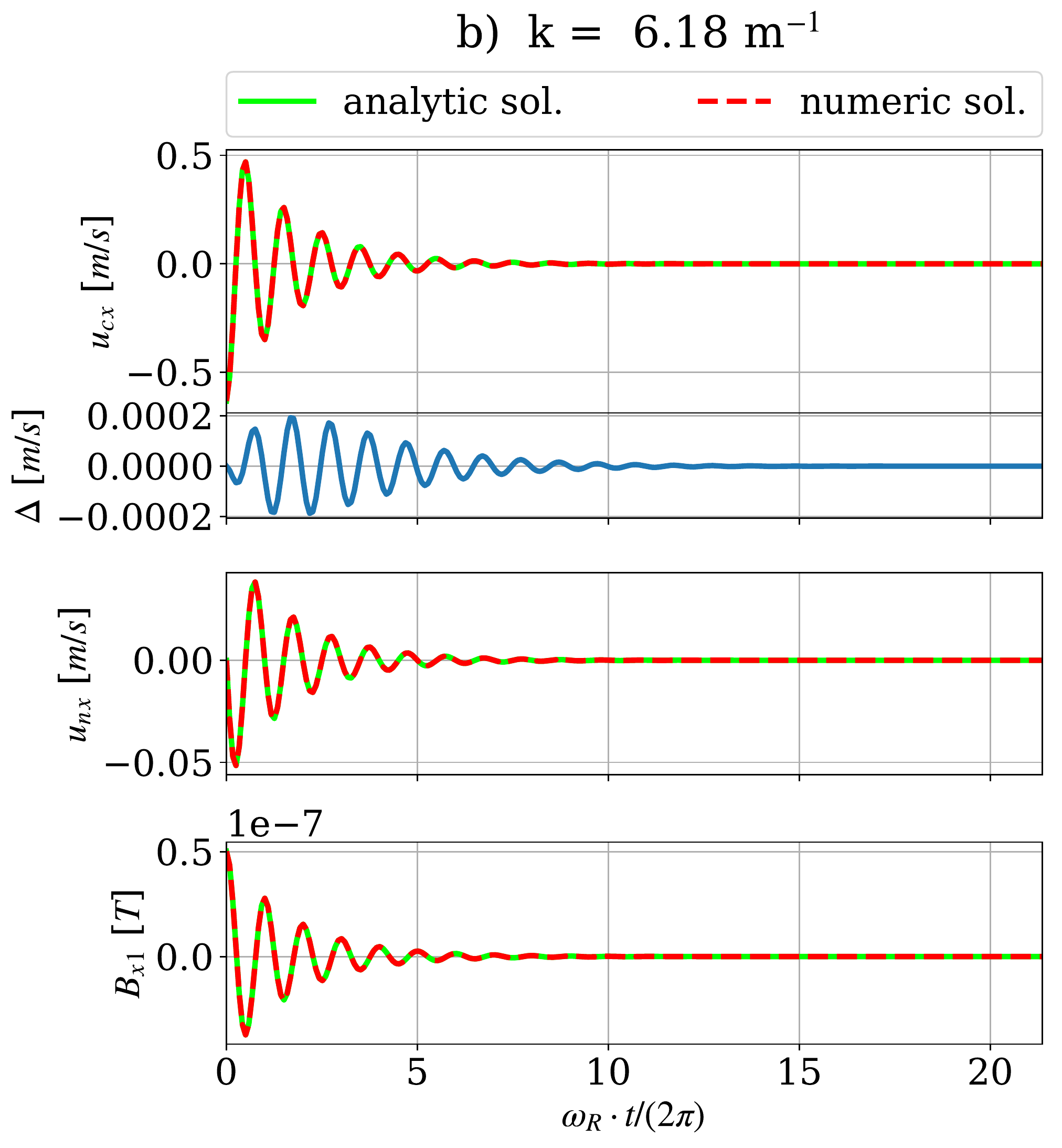}

\includegraphics[width = 8.5cm]{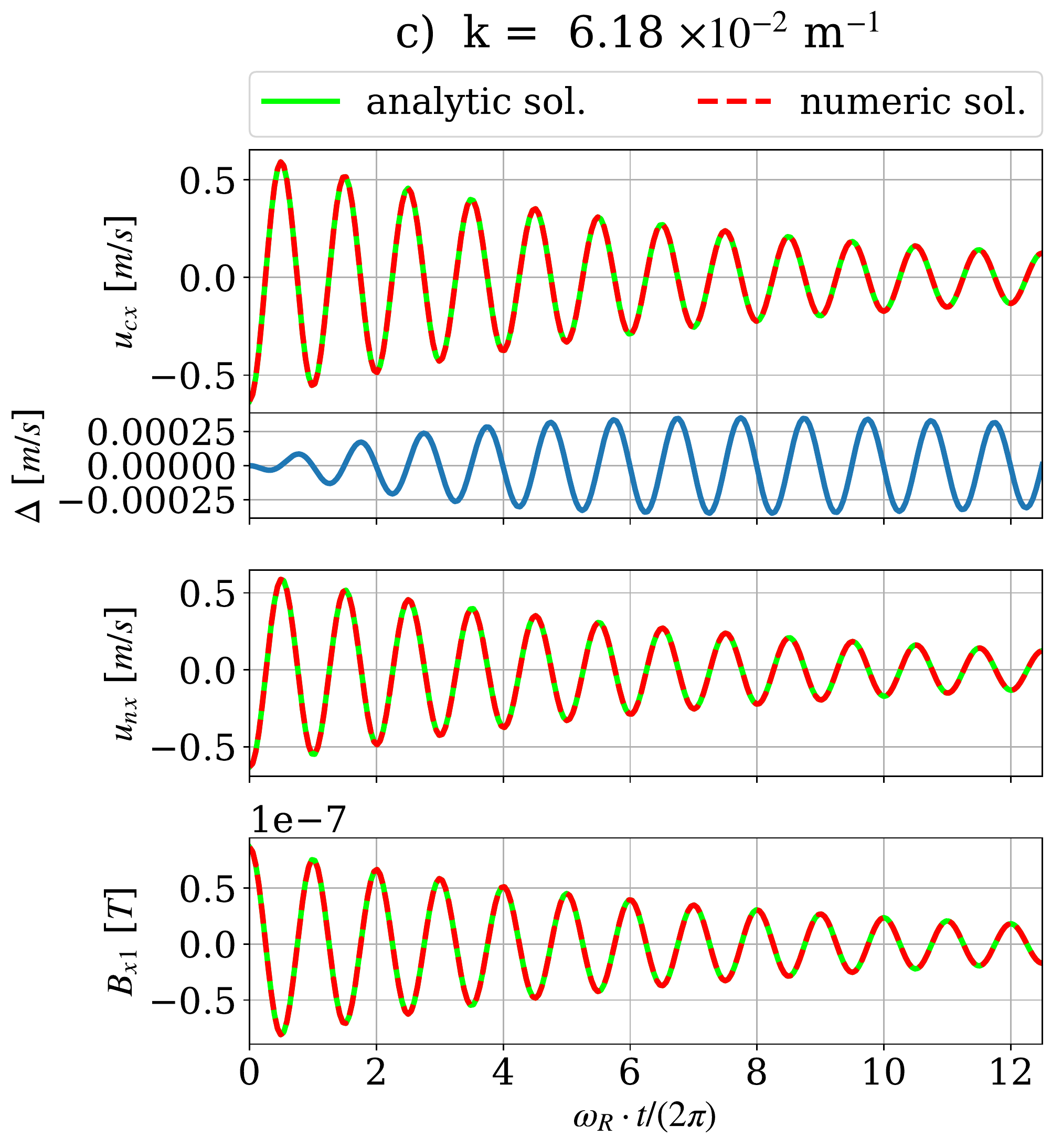}
\includegraphics[width = 8.5cm]{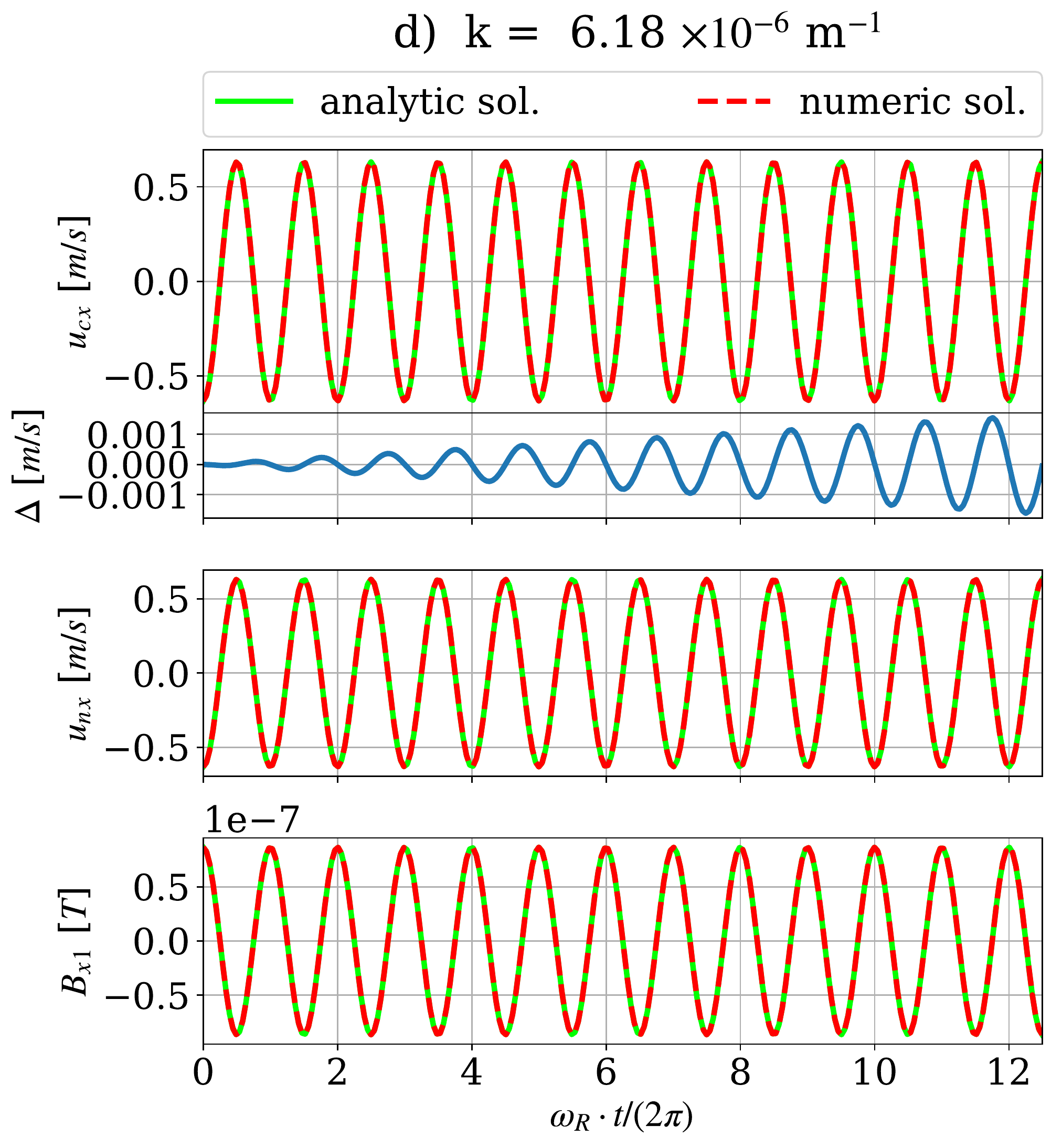}
\caption{Simulations of Alfv\'en waves propagation in a homogeneous plasma. Numerically (red, dashed line) and analytically (green, solid line) calculated time evolution of the velocity of neutrals and charges, and the magnetic field perturbation, as a function of time. 
Below the panels showing the velocity of charges, the difference ($\Delta$) between the numerical solution and the analytical solution for $u_{\rm cx}$ is plotted.
Time is measured in units of the wave period, $2\pi/\omega_R$. Panels from left to right and from top to bottom show simulations for different values of the wavenumber k, as indicated in the legend of Fig. \ref{fig:alfvenw}. }
\label{fig:alf_ti_ev}
\end{figure*}
%
\subsection{ Temporal damping of acoustic waves}  \label{tests_acoustic_time}
%
Figure \ref{fig:figAcT} shows the numerical and analytical solutions for the acoustic waves with wavenumber values of k = $618$, $6.18$, $2.06$, and $6.18 \times 10^{-6}$ m$^{-1}$.  

We can observe that for different values of k the amplitude damping and propagation velocity vary in agreement with what is expected from the Figure \ref{fig:acousticw}. For the large value of k = $618$ m$^{-1}$ (panel {\it a} of Figure \ref{fig:figAcT}) there is a phase shift between the  oscillating fluid velocities of neutral and charges, and both are slightly damped. Notice that the amplitude of oscillations of velocity of charges is significantly larger than that of the neutrals. For the intermediate values of  k = $6.18$ and $2.06$ m$^{-1}$ (panels {\it b} and {it c}  of Figure \ref{fig:figAcT}),  we observe significant damping of both velocity oscillations, but similar amplitudes and a smaller phase shift. For the small k case, k = $6.18 \times 10^{-6}$ m$^{-1}$ (panel {\it d})  the velocities of charges and neutrals are the same, and the wave propagation velocity is the sound speed of the plasma as a whole. This is expected since charges and neutrals are collisionally coupled. 

We observe from Figure \ref{fig:figAcT} that numerical and analytical solutions are in very good agreement. For all the panels in the Figure, we have plotted the difference between the  numerical and the analytical solution, labelled $\Delta$ right below the plots for $u_{\rm cx}$. It can be observed that the error accumulates in time,  but it is nevertheless very small.

The values from Table \ref{tab:charValAW} are consistent with the values deduced from Figure \ref{fig:acousticw}. The amplitude of a wave will decrease to  a fraction $f =  \text{exp}(-2 \pi P /T_D)$ in a period.  For example,  in the case of the wave with wavenumber k = 6.18 m$^{-1}$,  $f \approx \text{exp}(-9.6/7.37) \approx$ 0.27, and it is consistent with the value observed in panel {\it b} of Figure \ref{fig:figAcT}. 

\subsection{Temporal damping of Alfv\'en waves}

Figure \ref{fig:alf_ti_ev} shows the numerical and analytical solutions for the Alfv\'en waves with the wavenumber k  set to the following values,  k = $61.8$, $6.18$, $6.18 \times 10^{-2}$, and $6.18 \times 10^{-6}$ m$^{-1}$. These values of k are marked in  Figure \ref{fig:alfvenw}.

We observe that  numerical and analytical solutions match exactly and the amplitude damping and wave propagation speed vary with  the ratio $\alpha$/k as expected  from Figure \ref{fig:alfvenw}.

For small $\alpha/k$,  the wave propagates at the Alfv\'en speed of the charges and the velocity of neutrals is almost zero  (panel {\it a} of Figure \ref{fig:alf_ti_ev}).  For large $\alpha/k$,  the wave propagates at the Alfv\'en speed of the whole fluid and the neutrals and charges velocities are equal (panel {\it d}). 
Yet again, for intermediate values of  $\alpha/k$, the damping of all perturbations is observed (panels {\it b} and {\it c}).

For all the panels in the Figure, we show the difference between the  numerical and the analytical solution of $u_{\rm cx}$, labelled $\Delta$, right below the plots for $u_{\rm cx}$.  The error, similarly to the case of the acoustic waves, accumulates in time, and is small. For the Alfv\'en waves, as well, the values from Table \ref{tab:charValAlf} are consistent with the values obtained from the numerical simulations. For example, the wave with wavenumber k = 6.18 m$^{-1}$, should decrease the amplitude  to a fraction $f\approx \text{exp}(-1.63/2.76) \approx$ 0.55 in a period, and this value can also be deduced from panel {\it b} in Figure \ref{fig:alf_ti_ev}. 

\subsection{Time convergence test}

\begin{figure*}
\centering
\includegraphics[width = 6cm]{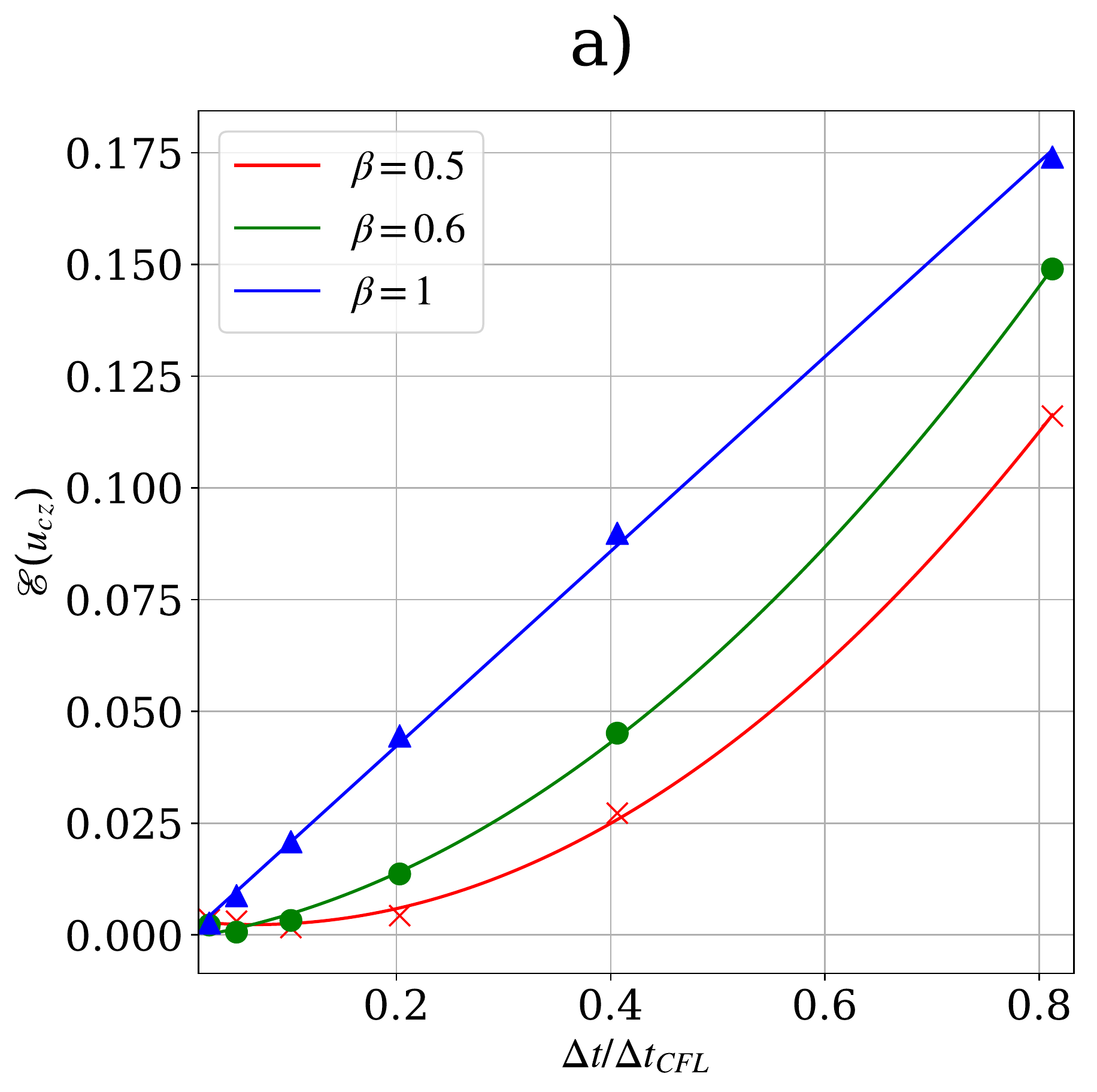}
\includegraphics[width = 6cm]{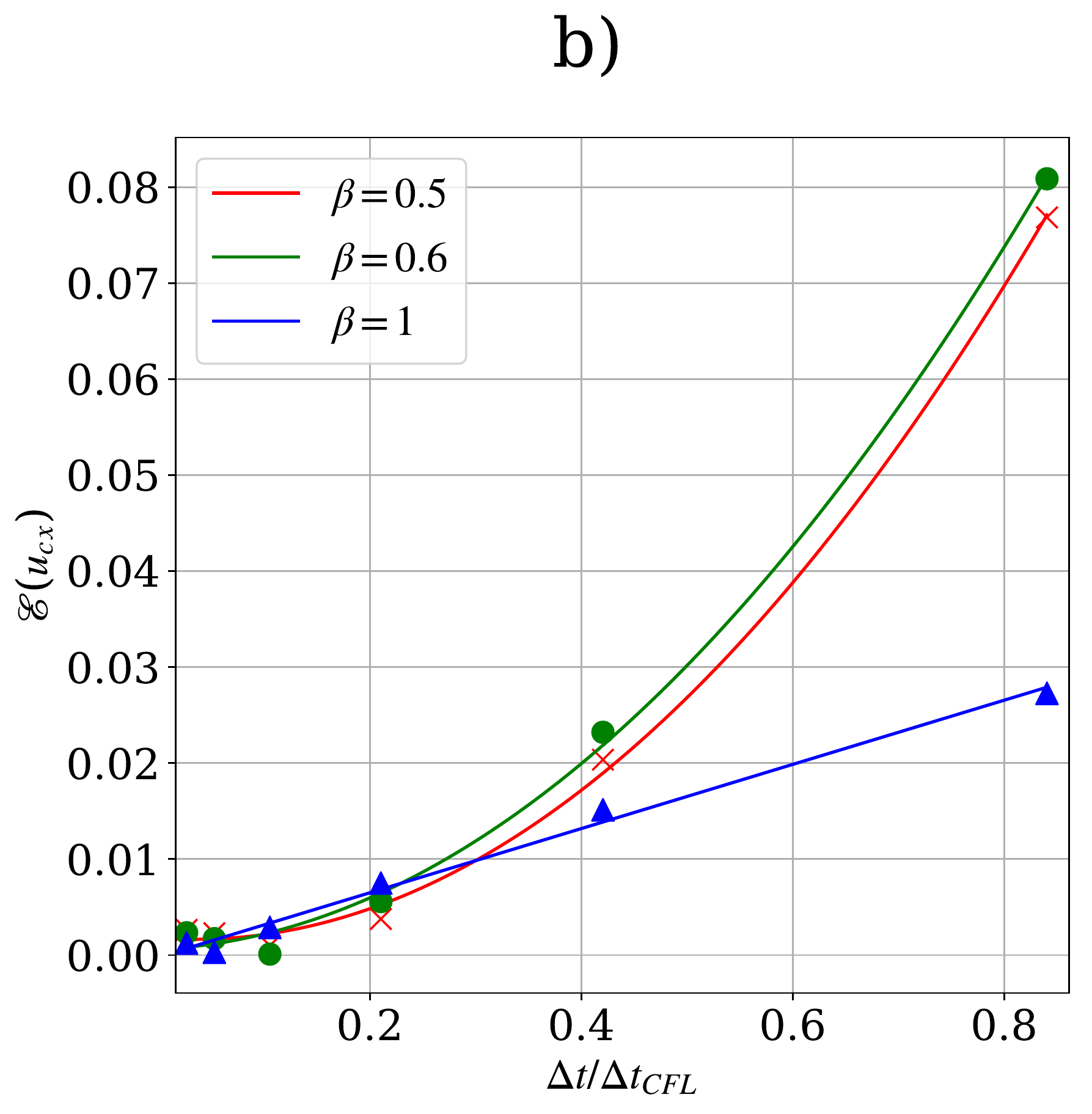}
\caption{Normalized errors, $\varepsilon$, defined by Eq. \ref{eq:error}, in the velocity of charges  as a function of the ratio between the integration time step $\Delta t$ and ${\Delta t}_{CFL}$, accumulated after running the simulation until t=$t_F$.Panel a) corresponds to simulations of acoustic waves, and panel b) to simulations of Alfv\'en waves. Blue triangles show results for $\beta=1$ (defined in Eq. \ref{eq:beta}); green circles are for $\beta=0.6$ and red crosses are for $\beta=0.5$. Solid lines show the results of quadratic polynomial fit $\varepsilon=a(\Delta t/{\Delta t}_{CFL})^2  + b (\Delta t/{\Delta t}_{CFL}) + c$, with coefficients given in Tables \ref{tab:coefFitConv1} and \ref{tab:coefFitConv2}.}
\label{fig:figConv}
\end{figure*}
%
In order to quantify the degree of agreement between the analytical and numerical solutions, we performed 
a time-step convergence test.  We study the normalized error, $\varepsilon(f,g)$, between analytical and numerical solutions as a function of the ratio between the integration time step of the simulations normalized, $\Delta t$ and the time step imposed by the single-fluid CFL condition, ${\Delta t}_{CFL}$.  The normalized error is defined as:
\begin{equation} \label{eq:error}		
\varepsilon(f,g) = \sqrt{\frac{\sum_i{ (f(x_i) - g(x_i))^2 }}{\sum_i{ (g(x_i))^2 }} }.
\end{equation}
In this expression, $f$ and $g$ are numerical and analytical solutions, respectively, computed at grid points $x_i$ with i = 1,..N. 
The analytical solution, $g$, is independent of $\Delta t$.  The normalized error defined this way was computed independently for all perturbed variables of the simulations. These are $u_{\rm cz}, u_{\rm nz}, \rho_{\rm c1}, \rho_{\rm n1}, p_{\rm c1}, p_{\rm n1}$ for the acoustic waves, and $u_{\rm nx}$, $u_{\rm cx}$ and $B_{\rm 1x}$ for the Alfv\'en waves. The results for different variables came to be very similar, therefore we only show the error for the velocity of charges, $u_{\rm cz}$ (acoustic waves) and  $u_{\rm cx}$ (Alfv\'en waves).  Computations for different wavenumber k also lead to very similar results.  In the tests below we used k = $6.18$ m$^{-1}$ for acoustic waves and k = $6.18 \times  10^{-2}$ m$^{-1}$  for Alfv\'en waves.

Figure  \ref{fig:figConv} shows the results of the convergence tests for both types of waves. The errors are global errors  computed after running simulations of acoustic and Alfv\'en waves until $t=t_F$, and the total time $t_F$ corresponds to 15.96 periods for the acoustic waves and to 12.55 periods for the Alfv\'en waves. The maximum time step, $\Delta t$, used  for the convergence test is close to the time step imposed by the single-fluid CFL condition. The numerical solutions were computed using time-steps $\Delta t$, having the ratio $\Delta t/{\Delta t}_{CFL}=$ 0.812, 0.406, 0.203, 0.102, 0.051, 0.025 for the acoustic waves and  $\Delta t/{\Delta t}_{CFL} =$ 0.84, 0.42, 0.21, 0.105, 0.053, 0.026 for the Alfv\'en waves.  The convergence tests were  carried out  for three values of the Toth's scheme parameter $\beta$ of 0.5, 0.6 and 1. 

Figure  \ref{fig:figConv} shows that, as expected, the error between the analytical and the numerical solutions increases  with the increase of the integration time step. However, this increase is different for different values of the parameter $\beta$.  This parameter, defined in Eq. \ref{eq:beta}  determines the fraction of the implicit contribution to the numerical solution.  For $\beta=1$ the scheme is expected to be 1st order accurate in time, and for $\beta=0.5$ it is expected to be second order accurate.  In order to test if this is the case in our implementation, we performed polynomial fit to the error function using the  expression $\varepsilon=a (\Delta t/{\Delta t}_{CFL})^2  + b (\Delta t/{\Delta t}_{CFL}) + c$, with the constraint a$\ge$0.  The coefficients resulting from the fit are given in Table \ref{tab:coefFitConv1} for the acoustic waves  and in Table \ref{tab:coefFitConv2} for the Alfv\'en waves.  The results confirm that the polynomial fit to the numerical values is dominantly 2nd order for $\beta \approx 0.5$ and  1st order for $\beta=1$. We also observe that while higher order convergence of numerical schemes is generally desirable,  the absolute value of the error for the $\beta = 1$ tests at a practical time-step  is of similar magnitude (acoustic waves) or smaller (Alfv\'{e}n waves) than for the $\beta \approx 0.5$ tests. 

It must be noted that the numerical solution here uses  $\Delta t$ up to the value dictated by the explicit single-fluid  CFL condition. Therefore, we conclude that our implementation allows to efficiently overcome the small time step limitations implied by the  stiff collisional terms in the two-fluid model.
%
\begin{table} [b]
\begin{center}
\caption{Values of the coefficients, together with their standard deviations  obtained after the polynomial fit $\varepsilon=a(\Delta t/{\Delta t}_{CFL})^2  + b (\Delta t /{\Delta t}_{CFL})+ c$ to the numerically obtained errors $\varepsilon$, for the convergence curves in Figure \ref{fig:figConv}, panel {\it a}, for acoustic waves.}
\begin{tabular}{c c c c }
\multicolumn{4}{l}{\bfseries Acoustic waves}\\	
\hline		
$\beta$ & a$/ 10^{-1}$    &  b$/ 10^{-2}$ & c$/ 10^{-3}$ \\
\hline			
0.5 &  $2.06 \pm 0.11$ & $-2.83 \pm 0.92 $     & $3.27 \pm 1.19$      \\
0.6 &  $1.82 \pm 0.12$   & $3.71 \pm 1.02$   & $-0.91 \pm 1.31$    \\
1    &  $0 \pm 0.16$      & $21.7 \pm 1.39$   & $-1.31 \pm 1.79$   \\
\hline  
\end{tabular}
\label{tab:coefFitConv1}
\end{center}
\end{table}
%
\begin{table}[t]
\begin{center}
\caption{Values of the coefficients, together with their standard deviations  obtained after the polynomial fit  $\varepsilon=a (\Delta t/{\Delta t}_{CFL})^2  + b (\Delta t/{\Delta t}_{CFL}) + c$ to the numerically obtained errors $\varepsilon$, for the convergence curves in Figure \ref{fig:figConv},  panel {\it b}, for Alfv\'en waves.}
\begin{tabular}{c c c c}
\multicolumn{4}{l}{\bfseries Alfv\'en waves}	\\
\hline			
$\beta$ & a$/10^{-1}$     & b$/10^{-3}$  & c$/10^{-3}$  \\
\hline			
0.5 & $1.16 \pm 0.10$              & $-7.85 \pm 9.02$  & $3.27 \pm 1.19$   \\ 
0.6 & $1.08 \pm 0.12$              & $ 5.38 \pm 10.9$  & $0.59 \pm 1.45 $ \\
1    & $0 \pm 0.08$                   & $ 33.4 \pm 7.26$  & $-0.18 \pm 0.97$ \\
\hline  
\end{tabular}
\label{tab:coefFitConv2}
\end{center}
\end{table}
%
\section{Waves in a gravitationally stratified atmosphere} \label{sec:test_grav}
In this section we test the capabilities of the code to model waves in a strongly gravitationally stratified solar chromosphere.
We assume a model atmosphere with all hydrodynamical parameters and purely horizontal magnetic field, $B_{\rm x0}$, stratified in the vertical, $z$, direction. 
The temperature is considered uniform  (height-independent), and different for charges and neutrals.
If we neglect viscosity, and consider only elastic collisions, adiabatic equation of energy, and ideal Ohm's law, the linearized
equations can be written as:

\begin{eqnarray} 
\frac{\partial \rho_{\rm c1}}{\partial t} &+& u_{\rm cz} \frac{d \rho_{\rm c0}}{d z} + \rho_{\rm c0}  \frac{\partial u_{\rm cz}}{\partial z} = 0, \nonumber  \\
\frac{\partial \rho_{\rm n1}}{\partial t} &+& u_{\rm nz} \frac{d \rho_{\rm n0}}{d z} + \rho_{\rm n0}  \frac{\partial u_{\rm nz}}{\partial z} = 0, \nonumber  \\
\rho_{\rm c0} \frac{\partial u_{\rm cz}}{\partial t} &+&\rho_{\rm c1} g + \frac{\partial p_{\rm c1}}{\partial z} + \frac{1}{\mu_0}\left ( \frac{\partial B_{\rm x1}}{\partial z} B_{\rm x0}
 + \frac{d B_{\rm x0}}{d z} B_{\rm x1}\right ) \nonumber \\
&=&  \alpha \rho_{\rm n0} \rho_{\rm c0} (u_{\rm nz} - u_{\rm cz} ), \nonumber \\
\rho_{\rm n0} \frac{\partial u_{\rm nz}}{\partial t} &+& \rho_{\rm n1} g + \frac{\partial p_{\rm n1}}{\partial z} =  \alpha \rho_{\rm n0} \rho_{\rm c0} 
(u_{\rm cz} - u_{\rm nz} ), \nonumber \\
\frac{\partial p_{\rm c1}}{\partial t} &-& c_{\rm c0}^2 \frac{\partial \rho_{\rm c1}}{\partial t} - c_{\rm c0}^2 u_{\rm cz} \frac{d \rho_{\rm c0}}{d z} + u_{\rm cz} \frac{d p_{\rm c0}}{d z} = 0, \nonumber \\
\frac{\partial p_{\rm n1}}{\partial t} &-& c_{\rm n0}^2 \frac{\partial \rho_{\rm n1}}{\partial t} - c_{\rm n0}^2 u_{\rm nz} \frac{d \rho_{\rm n0}}{d z} + u_{\rm nz} \frac{d p_{\rm n0}}{d z} = 0, \nonumber \\
\frac{\partial B_{\rm x1}}{\partial t}  &+& B_{\rm x0} \frac{\partial u_{\rm cz}}{\partial z} + u_{\rm cz} \frac{d B_{\rm x0} }{d z} = 0.
\end{eqnarray}
\noindent We separate the time dependence, assumed to be of form $\text{exp}(i \omega t)$, with constant $\omega$,  and combine the system into the equation of the vertical velocity for the charges:  $u_{\rm cz}(z,t) = \tilde u_{\rm cz}(z)\text{exp}(i \omega t)$,  obtaining a fourth order ODE:
\begin{eqnarray} \label{eq:comb2fl} 
\frac{{\rm d}^4 \tilde u_{\rm cz}}{{\rm d} z^4} a_c a_n   + \frac{{\rm d}^3 \tilde u_{\rm cz}}{{\rm d} z^3} \left (a_n b_c + a_c b_n \right ) + \nonumber \\
 \frac{{\rm d}^2 \tilde{ u_c}_z}{{\rm d} z^2} \left[ b_c b_n + \omega^2 \left (a_c + a_n\right ) - i \alpha \omega (a_c \rho_{\rm c0} + a_n \rho_{\rm n0}) \right ]  +\nonumber\\
\frac{{\rm d} \tilde u_{\rm cz}}{{\rm d} z} \omega \left[ \omega \left(b_c + b_n\right) - i \alpha \left(b_c \rho_{\rm c0} + b_n \rho_{\rm n0} \right)  \right] +\nonumber \\
u_{\rm cz} \omega^3 \left[\omega - i \alpha \left(\rho_{\rm c0} + \rho_{\rm n0}\right) \right ] =0 \,.
\end{eqnarray}
where:
\begin{align} \label{eq:Hz2fl}
a_c(z) &=  {c_{\rm c0}}^2 + {v_{A0}}^2,  a_n(z) =  c_{\rm n0}^2, 
b_{c,n}(z)   =  \frac{1}{\rho_{\rm {c,n}0}} \frac{\partial \left( {\rho_{\rm {c,n}0}} a_{c,n}   \right)}{\partial z},  \nonumber \\
c_{\rm {c,n}0}^2 &= \gamma \frac{p_{\rm {c,n}0}}{\rho_{\rm {c,n}0}}, \, \, \, \, \,\text{and} \, \, \, \, \, 
{v_{A0}}^2 = \frac{{B_{\rm x0}}^2}{\mu_0 \rho_{\rm c0}}. 
\end{align}
\noindent The equilibrium variables for neutrals and for charges must fulfill  the equations of state: Eqs (\ref{eq:eos}), and
the hydrostatic (HS) and the magneto-hydrostatic (MHS) equlibrium conditions, respectively, which are:
\begin{eqnarray*}
\frac{d p_{\rm n0}}{dz} + \rho_{\rm n0} g = 0, \\
\frac{d}{dz} \left ( p_{\rm c0} + {B_{\rm x0}}^2/(2\mu_0) \right ) + \rho_{\rm c0} g = 0.
\end{eqnarray*}
\noindent Since the temperature is assumed uniform, the pressure for neutrals has an exponential profile with a uniform scale height,  and the sound speed of neutrals  is constant. If we consider that the magnetic pressure has the same scale height as the charges pressure, after solving HS/MHS equations we obtain: 
\begin{eqnarray}
p_{\rm n0}(z) &=& p_{\rm n0}(z_0) {\exp}\left( -\frac{z}{H_n}  \right), \nonumber \\
p_{\rm c0}(z) &=& p_{\rm c0}(z_0) {\exp}\left( -\frac{z}{H_c}  \right), \nonumber \\
B_{\rm x0}(z) &=& B_{\rm x0}(z_0) {\exp}\left( -\frac{z}{2 H_c}  \right).
\end{eqnarray}
with uniform scale heights:
\begin{eqnarray}
H_n &=& \frac{k_B T_{n0}}{m_H g}, \nonumber \\ 
H_c &=& \frac{2 k_B T_{c0}}{m_H g } \frac{p_{\rm c0}(z_0) + B_{\rm x0}(z_0)^2/(2 \mu_0)}{p_{\rm c0}(z_0)}.
\end{eqnarray}
The densities obtained from the ideal gas laws for neutrals and charges, Eqs. \ref{eq:eos}, also have an exponential profile.
In these conditions, the quantities defined in Eq. \ref{eq:Hz2fl}:  $a_{c,n}$, $b_{c,n} = -a_{c,n}/ H_{c,n}$n and $\alpha$ are uniform, however, there are coefficients that explicitly contain density. We assumed that the scale of the height variation of the non-uniform coefficients in Eq. \ref{eq:comb2fl} is large compared to the oscillation wavelength, and therefore we can search for the solution in terms of plane waves, as: 
\begin{equation*} 
\tilde u_{\rm cz}(z) = V_c \text{exp}\left(-i k z\right),
\end{equation*} 
with $V_c$ being the uniform real amplitude, and k the uniform complex wavenumber.
\noindent The other variables  also have form of plane waves (including the time dependence):
\begin{align}
\left\{ u_{\rm nz}, \frac{\rho_{\rm c1}}{\rho_{\rm c0}}, \frac{\rho_{\rm n1}}{\rho_{\rm n0}}, \frac{p_{\rm c1}}{p_{\rm c0}}, \frac{p_{\rm n1}}{p_{\rm n0}}, 
\frac{B_{\rm x1}}{B_{\rm x0}} \right\} =&\{\tilde  V_n, \tilde R_c,  \tilde R_n, \tilde P_c,  \tilde P_n, \tilde B \} \times
\nonumber \\ & \exp{\left( i (\omega t - k z)  \right)  },
\end{align}
where $\tilde V_n, \tilde R_c, \tilde R_n, \tilde P_c, \tilde P_n, \tilde B $ are uniform complex amplitudes.
\noindent The following relations are obtained,
\begin{eqnarray}	\label{eq:relaw32Fl}
 \tilde V_n &=& V_c \left ( 1 + \frac{i(\omega^2  - i k b_c- k^2 a_c) }{\alpha \omega \rho_{\rm n0}} \right), \nonumber \\
\tilde R_{c,n}  &=& \frac{\tilde V_{c,n} }{ \omega}\left( k - i \frac{1}{H_{c,n}} \right),  \nonumber \\
\tilde P_{c,n}  &=& \frac{\tilde V_{c,n}  }{\omega} \left( k \gamma - i \frac{1}{H_{c,n}}  \right), \nonumber  \\
\tilde B &=& \frac{V_c }{\omega} \left( k - i \frac{1}{2 H_{c}} \right).	
\end{eqnarray}
\noindent The dispersion relation is a fourth order equation in $\omega$:
\begin{eqnarray}  \label{eq:dispRel2Fl}
(-\omega^2   + k^2 a_c  +i k b_c + i \omega \alpha \rho_{\rm n0}  ) \times \nonumber \\ 
(-\omega^2  + k^2 a_n + i k b_n  + i \omega \alpha \rho_{\rm c0} )+\omega^2 \alpha^2 \rho_{\rm n0} \rho_{\rm c0} = 0.
\end{eqnarray}
In our particular case, the wavenumber $k$ obtained from this space dependent dispersion relation results to be almost uniform, and therefore we consider the plane wave solution to be a fair approximation.

\subsection{Initial conditions}

\subsubsection{Equilibrium atmosphere}

We choose the temperature for neutrals: $T_{\rm n0}$ = 6000 K.  We use for the neutrals and charges  number density the values taken from the  VALC \citep{VALC} model at $z_0\approx 500$ km: $n_{\rm c0}(z_0)=5\times10^{17}$ m$^{-3}$, and  $n_{n0}(z_0) = 2.1\times10^{21}$ m$^{-3}$. In this test we take electrons into account, unlike the tests in the uniform atmosphere, they contribute to the pressure of charges (n$_{\rm c0} = 2 n_{\rm i0} = 2 n_{\rm e0}$), and to collisions, in
Equation (\ref{eq:alpha}). We choose the value of $B_{\rm x0}(z_0) = 10^{-4}$ T. In order to have the same scale height for neutrals and charges, we choose the temperature of the charges:
\begin{equation}
T_{\rm c0} = \frac{n_{\rm c0}(z_0) k_B T_{\rm n0} - B_{\rm x0}(z_0)^2/(2 \mu_0)}{2 n_{\rm c0}(z_0) k_B} 
\end{equation}
This gives the value of $T_{\rm c0}  \approx 2422.47$ K, and  $H_n = H_c \approx 1.8 \times 10^5$ m. Then, the pressures of charges and neutrals at $z_0$, $p_{\rm c0}(z_0)$ and $p_{\rm n0}(z_0)$,  are obtained from the ideal gal law, Eq. \ref{eq:eos}.  In these conditions, the wavenumber k obtained from the dispersion relation Eq. (\ref{eq:dispRel2Fl}) is almost uniform, 
and has a value $k\approx 1.383\times10^{-4} + 2.767 \times 10^{-6} i$ m$^{-1}$. Therefore, the wavelength is about 4 times shorter than the density scale height. The densities are calculated afterwards from the ideal gas laws for neutrals and charges, Eqs. \ref{eq:eos}, taking $\rho_{\rm n0}=n_n m_H$, $\rho_{\rm c0}=n_e m_H$. We cover the domain $L_z$ = 1.6 Mm with 32000 grid points.

\subsubsection{Perturbation}

We choose the period of the wave $P$=5 s, and calculate the frequency: $\omega = 2 \pi/P$, and the wavenumber, k from the dispersion relation Eq. (\ref{eq:dispRel2Fl}). We choose the  amplitude of the perturbation of the velocity of charges as a fraction of the background sound speed: $V_c = 10^{-3} c_0$,  where $c_0 = \sqrt{\gamma (p_{\rm n0} + p_{\rm c0})/(\rho_{\rm n0} + \rho_{\rm c0})}$ is the sound speed of the whole fluid, and its value is $c_0 \approx $ 9.1 km/s.  We calculate the  amplitudes of the other perturbations from the polarization relations, Eqs (\ref{eq:relaw32Fl}). The perturbation is generated by  a driver at each time step at the bottom of the atmosphere, in the ghosts points, which makes it the lower boundary condition. The Perfectly Matched Layer (PML) is used as the upper boundary condition. PML  is specially designed to absorb waves
without reflections. It was first introduced for the first time for electromagnetic waves in Maxwell equation by \citet{Berenger1994}, applied to Euler equations by \citet{Hu1996} and to acoustic waves in a strongly stratified solar convection zone by \citet{2007ManchaPa}. The description of the implementation of PML in Mancha code can be found in \cite{Felipe2010}. In this test we have used the value of the scheme parameter $\beta$=1.

\subsection{Results}

\begin{figure*}
\centering
\includegraphics[width = 12cm]{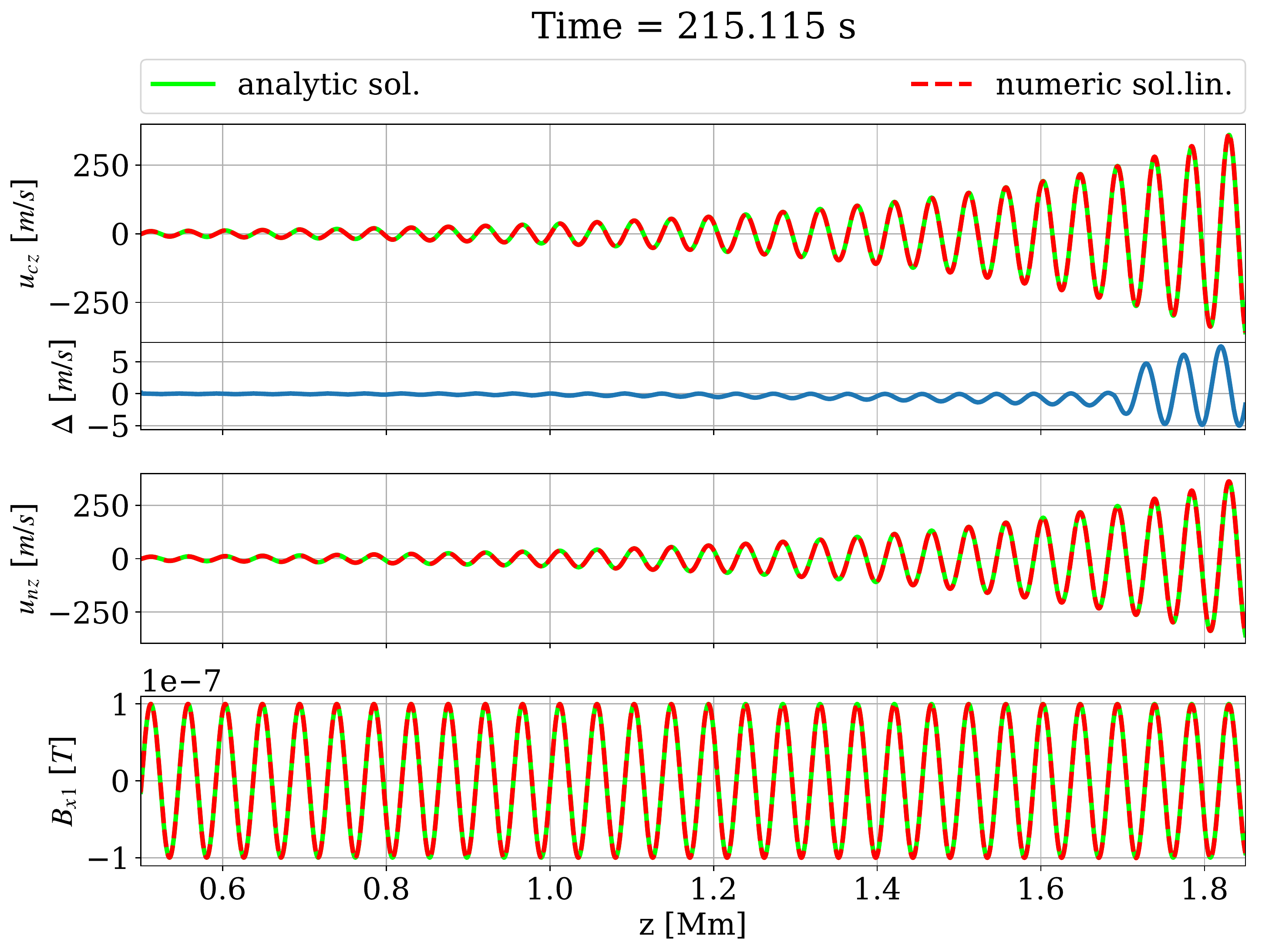}
\caption{Analytical (green, solid line) and  numerical linear(red, dashed line) for the vertical velocity of charges ($u_{\rm zc}$), and neutrals ($u_{\rm zn}$), and for the perturbation in the x component of the magnetic field ($B_{\rm 1x}$) for a snapshot taken in stationary state at time=215.115 s.  Below the panel showing the solutions for $u_{\rm zc}$, the difference between the numerical linear solution and the analytical solution for   $u_{\rm zc}$ is plotted.}
\label{fig:testNonuni}
\end{figure*}

\begin{figure*}
\centering
\includegraphics[width = 12cm]{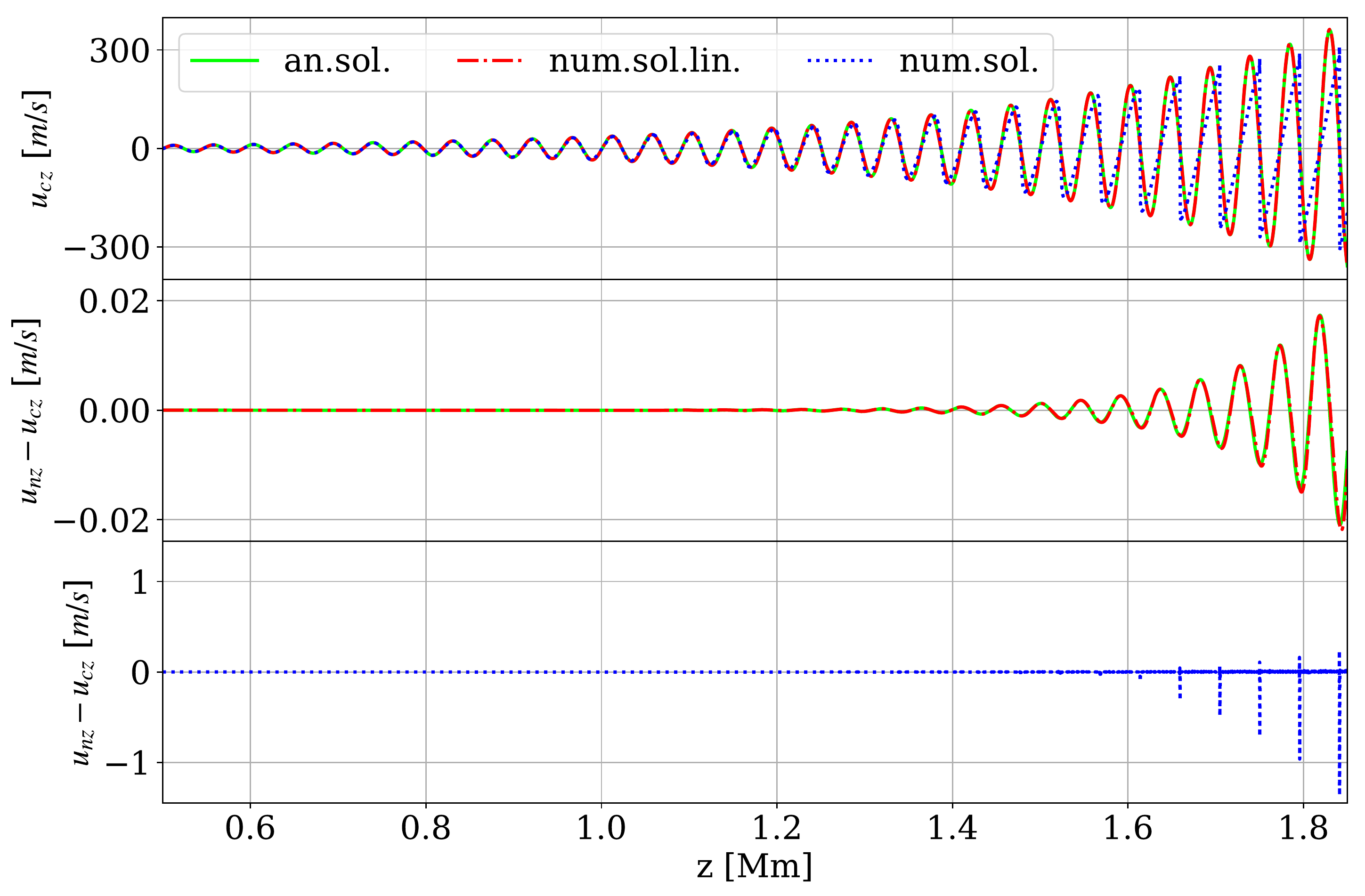}
\caption{First panel: Analytical solution (green, solid line), and numerical: linear (red, dasehd line) and nonlinear (blue, dotted line) for $u_{\rm zc}$ for a snapshot taken in stationary state at time=215.115 s. Second and third  panels: decoupling ($u_{\rm nz} - u_{\rm cz}$) for the analytical and linear numerical solutions, and  for the nonlinear numerical solutions respectively. }
\label{fig:testNonuni2}
\end{figure*}

 We run the code in two regimes. In the first case, we solved fully-nonlinear equations for perturbations, and in the second case we evolved linearized equations where only the first order terms were kept. The latter was done for comparison purposes, since the analytical solution assumes linear regime.  We compare  the numerical solutions at time t=215.115 s with the analytical solution. At this time simulations reached the stationary state, since the wave has reached the upper boundary and several periods of the wave have passed through the boundary. The results are shown  in Figure \ref{fig:testNonuni} and Figure \ref{fig:testNonuni2}.

Figure \ref{fig:testNonuni} shows the analytical solution (green, solid line)  superposed on the linear numerical solution (red, dashed line) for the vertical velocity of charges ($u_{\rm zc}$) in the first  panel, the vertical velocity of neutrals ($u_{\rm zn}$) in the second  panel,  and for the perturbation in the x component of the magnetic field ($B_{\rm x1}$) in the third  panel. Below the  panel of $u_{\rm zc}$ we show the difference ($\Delta$) between the linear numerical and the analytical solution of $u_{\rm zc}$. We  observe that the numerical linear solution is in very good agreement with the analytical solution for the three quantities considered, and that the error is small (below 2\%).  

In Figure \ref{fig:testNonuni2} we show  in the first  panel the nonlinear effects by plotting, besides  the analytical solution (green, solid line) and  the linear numerical solution (red, dashed line),  the nonlinear numerical solution (blue, dotted line) of $u_{\rm zc}$, superposed, for the same snapshot taken at t=215.115 s. In the second and third  panels we show the decoupling in the vertical velocity ($u_{\rm nz} - u_{\rm cz}$)  for the analytical and linear solution superposed, 
and the nonlinear solution, respectively.

We  observe that the wave  profile steepens at the end of the domain, when the amplitude becomes large,  and nonlinear effects are visible in the case of the nonlinear solution. As a consequence, the amplitude of the wave is smaller than in the linear case. 
\citep[see e.g.][]{Landau1987Fluid}. Wave amplitude grows with height in a gravitationally stratified atmosphere because of the density decrease. In the case considered here,  the damping is not large enough to overcome this growth, therefore the wave evolves into a shock. The amplitude growth is nevertheless below the growth in the MHD limit.

\begin{figure}
\centering
\includegraphics[width = 8.5cm]{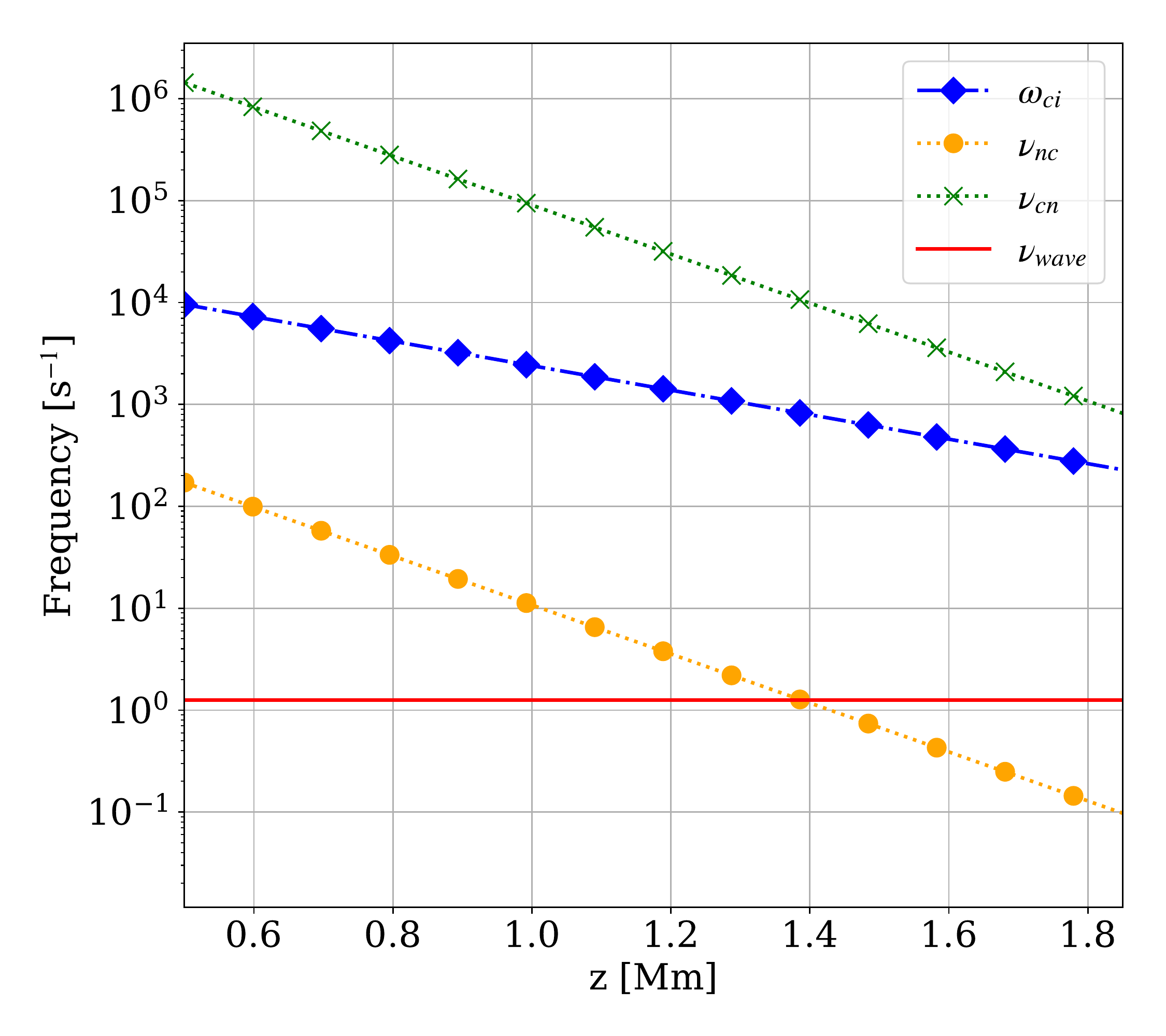}
\caption{Relevant frequencies: ion-cyclotron frequency: $\omega_{ci}$ (blue, dashed line, marked with diamonds), wave frequency: $\nu_{wave}$ = $2 \pi / {P}_{wave}$ (red, solid line), neutral-charge collision frequency: $\nu_{nc}$ (orange, dotted line, marked with circles), and charge-neutral collision frequency: $\nu_{cn}$ (green, dotted line, marked with "X")  }
\label{fig:freqNonuni}
\end{figure}

We  also observe that the decoupling predicted analytically  from the relation between $\tilde V_n$ and $V_c$ in Equation (\ref{eq:relaw32Fl}) agrees with the decoupling obtained from the linear numerical simulation. In order to understand intuitively the reason for the linear decoupling, we show in Figure \ref{fig:freqNonuni} the relevant frequencies corresponding to this problem. Even though the collisional parameter $\alpha$ is uniform because of the uniform background temperature of neutrals and charges, and has the value $\approx 4.1 \times 10^{11} m^3/kg/s$, the collision  frequency also depends on density (Eqs. \ref{eq:eff_coll}),  and has an exponential profile.  While for the charges, the collision frequency $\nu_{cn}$ is larger than both the ion-cyclotron frequency ($\omega_{ci}$) and the wave frequency (which determine the hydrodynamical time scale), for the neutrals, the neutral-charges collision frequency goes from being greater than  the wave frequency to being less than the wave  frequency at a point located at z $\approx$ 1.4 Mm in the atmosphere. This is the point after which we  observe the decoupling for the linear numerical solution and the analytical solution. The code captures well the transition from a coupled  to a partially decoupled regime. In the nonlinear case the hydrodynamical time-scale is  smaller than in the linear case.  The shock front width which determines the  hydrodynamical space scale in the nonlinear case is smaller than the wavelength. The nonlinear decoupling appears spatially at the shock front, and is almost five times larger than the linear decoupling.

%

\section{Discussion and conclusions}
%
Strong vertical and horizontal stratification in solar atmospheric plasma parameters makes particularly complicated theoretical modeling of the solar chromosphere, where it is expected that the collisional time scales between charged and neutral plasma components may become similar or longer than hydrodynamical time scale, leading to a breakdown of the single-fluid MHD approximation. In such conditions, neutrals start to decouple from charges, behaving as two independent fluids. This work describes the results of our numerical effort to extend the existing single-fluid non-linear MHD code  \mancha for modeling of solar plasma dynamics under two-fluid approximation. To this end, we have implemented a semi-implicit numerical scheme following the approach by \citet{2012Toth}.  Such development is a necessary logical step towards creation of a 3D  modeling tool for simulations  of multi-component plasma processes in the solar chromosphere.

As discussed in the Introduction, the collisional terms in the upper layers of the solar atmosphere have values similar to the rest of the forces in the momentum equation. In the lower layers (photosphere) their values significantly dominate over the rest of the terms. However, coupling between these layers plays an important role in the energy and momentum transfer from the solar interior to the corona, and therefore, it is necessary to create numerical tools that would be able to treat both extreme situations in a single simulation domain. Classical explicit schemes would be limited by the explicit time step dictated by the CFL condition due to collisional terms, making it very small and slowing significantly the numerical code. A suitable alternative is a semi-implicit implementation, suggested in \citet{2012Toth}. This implementation allowed us to keep the advantages of an explicit code (as efficient parallelization, lower memory requirements), but at the same time to overcome the restrictions imposed by small collisional times. We based our implementation on already existing numerical code, \mancha, 
thus being able to keep all its already implemented functionalities such as hyper-diffusivities and noise filtering modules, Perfectly Matched Layer boundary condition, and parallelization. 

Several features of our numerical code,  such as shock resolving,  have been previously thoroughly tested  in the past\citep{2006ManchaKh, Felipe2010, Khomenko+Collados2012, Pedro2018}. The particular effort in this work was to verify the performance of the semi-implicit fluid coupling algorithm.  For that we have compared the numerical solution with the known analytical solutions for propagation of acoustic and Alfv\'en waves in a homogeneous plasma with a different degree of collisional coupling. 

Our analytical results are similar to those obtained by \cite{2011Zaq} and \citet{2013Soler}.  When the temperatures of charges and neutrals are initially different, if the  ratio $\alpha$/k is small, the acoustic waves  in each of the fluids propagate nearly independently of each other with their respective sound speeds.  The Alfv\'en waves propagate with the Alfv\'en speed of the charged fluid if the collisional coupling is weak.  In the opposite limit, when the ratio $\alpha$/k is large, the two fluids become coupled.  The acoustic waves propagate with the sound speed of the whole fluid, and the Alfv\'en waves with the Alfv\'en speed  calculated using the total atom mass density.  In this case, the velocities of neutrals and charges are of equal amplitude and in phase.  At these two extremes, when the quantity $\alpha \rho_{\rm tot}$/(k $c_{\rm tot}$)  is either low or high, the waves have little damping relative to wavenumber,  but when the collisional frequency is of order of the wave frequency the damping  relative to wavenumber is maximum.

Our numerical solution  reproduces the analytical solution for the full range of the values of the wavenumber k. The temporal convergence tests have shown that, as expected, the newly implemented scheme is 1st other accurate when the scheme parameter $\beta=1$,  and is second order accurate when  $\beta \approx 0.5$.  However, we note that even if the scheme converges linearly with $\Delta t$  for the scheme parameter $\beta=1$, the overall behavior of the scheme is more stable, and the errors are in fact  smaller.

 Our current study  must be viewed in the context of other similar developments by other groups of authors. 
There have been several approaches proposed to overcome the limitations introduced by large collisional terms. 

 \cite{2016Hillier} noticed that, since collisional terms introduce a very different time scale, the solution can be obtained semi-analytically. In their approach, the equations with only collisional terms are solved analytically and then the rest of the terms are evolved numerically using an explicit scheme. When the hydrodynamical and the collisional time scales are similar, only explicit scheme is applied. Such an approach has a drawback since the entire numerical domain has to be placed in one or another regime, and therefore it may cause problems when strong stratification in the atmosphere is present. 
In our case, there is no need to have a criterion to distinguish between these regimes, the scheme handles well both of them, in the same way. 
The convergence tests were  carried out for the coupled regime, where it shown that the errors are larger. 
The neutrals and charges are more coupled at larger values of the ratio $\alpha/k$.
We have seen from the convergence test, that the scheme behaves better in the strongly coupled regime for the values
of the scheme parameter $\beta$ closer to 1 than to 0.5, and this is the reason for choosing this value for the last test. The last test, the simulation in the gravitationally  stratified stmosphere,
 deals with the situation  where the waves passes from a collisionally coupled 
regime to a decoupled regime, and the numerical results are satisfactory.

Alternatively, \cite{2017Maneva} used a fully implicit scheme which has no restriction for the time-step and is generally more stable. It must be noted that despite the advantage of the stability, implicit schemes are more expensive computationally, because, generally, the matrix inversions needed in the implicit part are usually implemented by iterations, increasing the computational time. However, if one only deals with the collisional terms as $S_n$, $\vec{R}_n$ and $M_n$ in equations \ref{eq:s}, \ref{eq:r} and \ref{eq:m}, there is an important advantage since these terms are linear with respect to the main set of variables and do not contain derivatives. Therefore the matrix inversion can be done fully analytically, improving the precision and the computational capabilities of the implicit code. Such analytical approach is not possible when dealing with physical effects such as thermal conduction or viscosity. In general, semi-implicit 
implementations such as the one presented here are less computationally expensive, but also less accurate, than fully implicit ones. 

Similarly to  our method, \citet{2008Sakai} have implemented the collisional terms following a semi-implicit approach.\citet{2008Sakai} use a two fluid code, where they consider the plasma  composed by ions and neutrals, and they implement most of the collisional terms in a scheme similar to Toth's scheme with $\beta$ parameter equal to 1. These authors considered ionization/recombination terms in continuity and momentum equations, elastic collisions in momentum equations,  and the work done by the collisional terms in the energy equations. 
However, unlike our implementation, they did not consider the inelastic collision terms, the thermal exchange, and  the frictional heating in the energy equations.

Yet another widely used code for multi-fluid simulations is  the HiFi code by \cite{2016Lukin},  which uses an implicit temporal discretization and spectral element spatial decomposition. This code has been extensively used for simulations of reconnection in partially ionized plasma {\citep{2012Lukin, 2013Lukin, Ni2018} and has similar advantages and drawbacks of the implicit implementation discussed above. 

 All in all, the above implementations have their advantages and disadvantages. In our case, we have chosen a semi-implicit approach because
the collisional terms that need to be implemented implicitly, 
are linear in the variables that evolve in time, and the implicit solve can be done
analytically. The tests we have  carried out in order to verify the code have shown satisfactory results.

While the present paper only presents  calculations in cases where analytical solutions exists and can be compared to the numerical solution, our future work described in Paper II \citep{Popescu2018b} demonstrates that our newly implemented algorithm is able to efficiently deal with chromospheric gravitational stratification and to be used for modeling of propagation of chromospheric shock waves under the two-fluid framework.
\begin{acknowledgements}
This work was supported by the Spanish Ministry of Science through the project AYA2014-55078-P and the National Science Foundation. 
It contributes to the deliverable identified in FP7 European Research Council grant agreement ERC-2017-CoG771310-PI2FA for the project "Partial Ionization: Two-fluid Approach".
The author(s) wish to acknowledge the contribution of Teide High-Performance Computing facilities to the results of this research. TeideHPC facilities are provided by the Instituto Tecnol\'ogico y de Energ\'ias Renovables (ITER, SA). URL: http://teidehpc.iter.es
\end{acknowledgements}
\bibliographystyle{aa}

\begin{appendix} 
\section{Expressions of terms} \label{app:collTerms}

\subsection{Collisions}

Expressions for $\Gamma^{\rm ion}$ and $\Gamma^{\rm rec} $as functions of $n_e$ and $T_e$ are given in \cite{1997Voronov} and \cite{2003Smirnov}:

\begin{equation} 
\Gamma^{\rm rec}  \approx \frac{n_e}{\sqrt{T_e^*}}	2.6 \cdot 10^{-19}; \,\,\,\, {\rm s^{-1}}
\end{equation}

\begin{equation} 
\Gamma^{\rm ion}  \approx n_e A \frac{1}{X + \phi_{\rm ion}/{T_e^*}}\left(\frac{\phi_{\rm ion}}{T_e^*}\right)^K  e^{-\phi_{\rm ion}/T_e^*}; \,\,\,\,  {\rm s^{-1}}
\end{equation}
where 
$\phi_{\rm ion} = 13.6 eV$, 
$T_e^*$ is electron temperature in eV, 
$A = 2.91 \cdot 10^{-14}$, 
$K$ = 0.39, and $X$ = 0.232

\noindent The elastic collisional frequency between particles of specie $\alpha$ with particles of specie $\beta$ is expressed as
\citep[see e.g.,][]{1965Braginskii}:
\begin{equation} \label{eq:nu_el} 
 \nu_{\alpha\beta}= n_\beta \frac{m_\beta}{m_\alpha + m_\beta } \sqrt{\frac{8 k_B T_{\alpha\beta}}{\pi m_{\alpha\beta } }} \Sigma_{\alpha\beta} 
\end{equation}
\noindent Notice that
$\nu_{\alpha\beta} \neq \nu_{\beta\alpha} $, but $\rho_\alpha \nu_{\alpha\beta} = \rho_\beta\nu_{\beta\alpha}$. In the expression for collisional frequency, $T_{\alpha\beta}= (T_\alpha +T_\beta)/2$ is the average temperature,  and $m_{\alpha\beta } = m_\alpha m_\beta/(m_\alpha + m_\beta ) $ is the reduced mass of particles $\alpha$ and $\beta$.
\noindent The elastic collisional parameter $\alpha$ defined in Equation (\ref{eq:alpha}) has the form:
\begin{equation} \label{eq:alpha_el} 
\alpha = \frac{m_{in}}{{m_n}^2} \sqrt{ \frac{8 k_B T_{cn}}{\pi m_{in}}}  \Sigma_{in} + \frac{m_{en}}{{m_n}^2} \sqrt{ \frac{8 k_B T_{cn}}{\pi m_{en}}}  \Sigma_{en}
\end{equation}

\noindent The charge-exchange (elastic) collisional parameter 
between particles of specie $\alpha$ (ions) and particles
of specie $\beta$ (neutrals), is approximately expressed as \citep[see, e.g.][]{2012Meier}:
\begin{equation}  \label{eq:alpha_cx}
\alpha_{cx} = \frac{1}{m_H}\left (V^{cx}_0 + V^{cx}_{1\alpha} + V^{cx}_{1\beta} \right) \Sigma_{cx}
\end{equation}
\noindent with:
\begin{eqnarray}
V^{cx}_0 &=& \sqrt{\frac{4}{\pi} {{v_T}_\alpha}^2 + \frac{4}{\pi} {{v_T}_\beta}^2   + {{v_{\rm rel}}_{\alpha\beta}}^2} \nonumber \\
V^{cx}_{1\alpha} &=& {{v_T}_\alpha}^2  \left( 4 (\frac{4}{\pi} {{v_T}_\beta}^2 + {{v_{\rm rel}}_{\alpha\beta}}^2 ) +
\frac{9 \pi}{4} {{v_T}_\alpha}^2	  \right)^{-0.5}
\nonumber \\
\Sigma_{cx} &=& 1.12 \times 10^{-18} - 7.15 \times 10^{-20} \text{ln} (V^{cx}_0)
\end{eqnarray}
\noindent where ${v_T}_\alpha$, and ${v_{\rm rel}}_{\alpha\beta}$ are the thermal velocity of particles of specie $\alpha$, 
and the module of the relative velocity between particles of specie $\alpha$ and particles of specie $\beta$, respectively:
\begin{eqnarray}
{v_T}_\alpha &=& \sqrt{\frac{2 k_B T_\alpha}{m_\alpha}}  \nonumber \\
{v_{\rm rel}}_{\alpha\beta} &=& \mid \vec{u_\alpha} -  \vec{u_\beta} \mid
\end{eqnarray}
\noindent If charge-exchange reactions are taken into account, the effective elastic collisional parameter is:
\begin{equation}  
\alpha_{eff} = \alpha + \alpha_{cx}
\end{equation}

\subsection{Viscosities and thermal conductivities} 

\citep[derived by][]{1965Braginskii}:

\begin{eqnarray}
\xi_\alpha &=& \frac{n_\alpha k_B T_\alpha}{\nu_{\alpha\alpha}} \nonumber \\
K_\alpha &=& \frac{4 n_\alpha k_B T_\alpha}{\nu_{\alpha\alpha} m_\alpha} 
\end{eqnarray}

\noindent Replacing for the collisional frequency, the above equations can be written as:

\begin{eqnarray} \label{eq:visc_th_coef}
\xi_\alpha &=& \frac{\sqrt{\pi k_B T_\alpha  m_\alpha}}{4 \Sigma_{\alpha\alpha}} \nonumber \\
K_\alpha &=& \sqrt{ \frac{\pi k_B T_\alpha}{m_\alpha}}\frac{1}{4 \Sigma_{\alpha\alpha}} 
\end{eqnarray}
\noindent This makes the viscosity and thermal conduction coefficient to depend only on the temperature.
For the viscosity and thermal conductivity of the charges, only the ions are taken into account.

%
\noindent The collisional cross sections used were:
$\Sigma_{nn} = 2.1 \times 10^{-18} \text{m}^{2}$, 
$\Sigma_{ei} = \Sigma_{ii} = \frac{40\pi}{3} \left ( \frac{e^2}{4\pi\epsilon_0 k_B T_c} \right )^2$,
where $e$ is the elementary charge, and $\epsilon_0$ the permitivity of free space, 
$\Sigma_{in} = 1.16 \times 10^{-18} \text{m}^{2}$, and $\Sigma_{en} =   10^{-19} \text{m}^{2}$.

\noindent $m_n$ = $m_H$ is the mass of the hydrogen atom.

\section{Analytical expressions of the Jacobian} \label{app:jac}

The analytical expressions of the Jacobian where we use the superscripts D, M, E for the continuity, momentum and energy equations, respectively,  are:

\subsection{Continuity equations}

\begin{equation} 
 \hat{J}^D = 
  \begin{pmatrix}
    - \Gamma_{ion}  & \Gamma_{rec} \\
    \Gamma_{ion}  & -\Gamma_{rec} 
  \end{pmatrix}  
\end{equation}

\subsection{Momentum equations}
 
\begin{equation}
 \hat{J}^M = 
  \begin{pmatrix}
    - \Gamma_{ion} - \alpha \rho_c & \Gamma_{rec} + \alpha \rho_n\\
    \Gamma_{ion} + \alpha \rho_c & -\Gamma_{rec} - \alpha \rho_n
  \end{pmatrix}  
\end{equation}

\subsection{Energy equations}

\begin{equation} 
\hat{J}^E = 
  \begin{pmatrix}
-\Gamma_{ion} -  \rho_c \alpha   &  \frac{1}{A}(\Gamma_{rec} + \rho_n \alpha )\\
\Gamma_{ion} +  \rho_c \alpha   & -\frac{1}{A}(\Gamma_{rec} +    \rho_n \alpha)
  \end{pmatrix}  
\end{equation}
where A=2 when charges=ions+electrons, and A=1 when electrons are not taken into account.
\begin{equation} 
 \hat{K} =  \hat{J}^M  
\end{equation}

\end{appendix}

\end{document}